\documentclass[sigconf, nonacm]{acmart}
\usepackage{natbib}
\usepackage{booktabs} 
\usepackage{float}
\usepackage{amsmath}
\usepackage{wrapfig}
\usepackage{placeins}
\usepackage{multirow}
\usepackage{array}
\usepackage{tabularx}
\usepackage{float}

\title{AI persuading AI vs AI persuading Humans: LLMs' Differential Effectiveness in Promoting Pro-Environmental Behavior \\ \vspace{0.2cm} \large{\textbf{Preprint}}}

\author{Alexander Doudkin}
\authornote{Both authors contributed equally to this research.}
\email{doudkin@mit.edu}
\affiliation{
  \institution{MIT Media Lab}
  \institution{Massachusetts Institute of Technology (MIT)}
  \city{Cambridge}
  \state{MA}
  \country{USA}
}

\author{Pat Pataranutaporn}
\authornotemark[1]
\email{patppat@mit.edu}
\affiliation{
  \institution{MIT Media Lab}
  \institution{Massachusetts Institute of Technology (MIT)}
  \city{Cambridge}
  \state{MA}
  \country{USA}
}

\author{Pattie Maes}
\email{pattie@media.mit.edu}
\affiliation{
  \institution{MIT Media Lab}
  \institution{Massachusetts Institute of Technology (MIT)}
  \city{Cambridge}
  \state{MA}
  \country{USA}
}

 \begin{teaserfigure}
    \centering
    \includegraphics[width=1\linewidth]{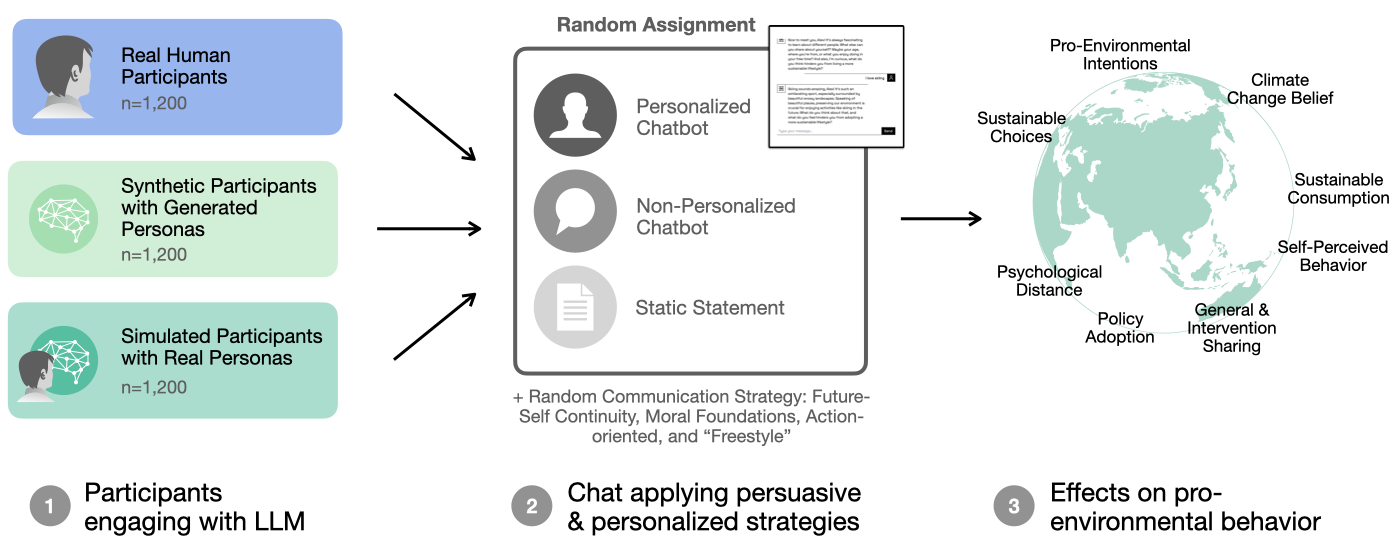}
    \caption{Overview of the system and research approach.}
    \Description{Teaser figure illustrating an overview of the system architecture and research approach for evaluating AI‐driven persuasive interventions for pro-environmental behavior. The schematic depicts key components including participant groups, intervention conditions, and the overall study flow.}
    \label{fig:title-system}
\end{teaserfigure}

\begin{abstract}
Pro-environmental behavior (PEB) is vital to combat climate change, yet turning awareness into intention and action remains elusive. We explore large language models (LLMs) as tools to promote PEB, comparing their impact across 3,200 participants: real humans (n=1,200), simulated humans based on actual participant data (n=1,200), and fully synthetic personas (n=1,200). All three participant groups faced personalized or standard chatbots, or static statements, employing four persuasion strategies (moral foundations, future self-continuity, action orientation, or ''freestyle'' chosen by the LLM). Results reveal a ''synthetic persuasion paradox'': synthetic and simulated agents significantly affect their post-intervention PEB stance, while human responses barely shift. Simulated participants better approximate human trends but still overestimate effects. This disconnect underscores LLM’s potential for pre-evaluating PEB interventions but warns of its limits in predicting real-world behavior. We call for refined synthetic modeling and sustained and extended human trials to align conversational AI’s promise with tangible sustainability outcomes.
\end{abstract}


\setcopyright{none}
\renewcommand\footnotetextcopyrightpermission[1]{}
\renewcommand\footnotetextcopyrightpermission[1]{} 
\begin{document}

\maketitle

\section{Introduction}
Climate change has become the quintessential challenge of our time, imperiling ecosystems, societies, and the broader stability of Earth’s support systems \cite{onClimateChangeIPCC2023ClimateChange}. Growing evidence and scientific consensus highlight the urgency of this crisis, yet translating environmental awareness into sustained action on an individual level remains elusive \cite{Myers2021ConsensusLater, Knutti2019ClosingChange, Lynas2021GreaterLiterature}. Understanding the drivers behind individuals' pro-environmental behavior (PEB) is a longstanding endeavor in psychology and social sciences, with prominent theoretical frameworks such as the Theory of Planned Behavior and meta-analytic studies on PEB predictors guiding the research \cite{Gifford2011TheAdaptation, Gifford2011BehavioralInterventions, Hines1987AnalysisMeta-Analysis, Bamberg2007TwentyBehaviour, Ajzen1991TheBehavior}. On an individual level, partially within policy context, many intervention and persuasion strategies---such as Nudging, leveraging norms, or future self-continuity---have shown significant influence on PEB \cite{Goldstein2008AHotels., Lehner2016NudgingBehaviour, Demarque2015NudgingEnvironment, Vlasceanu2024AddressingCountries, Duarte2024EnhancingInsights}. However, the so-called intention-action gap remains one of the key limitations for most of the interventions and strategies and their long-term efficacy \cite{Knutti2019ClosingChange, Nguyen2019GreenGap, Kollmuss2002MindBehavior}.
  
As a parallel development, innovative possibilities for encouraging human behavior have emerged through advances in conversational AI. Large language models (LLMs) enable real-time, adaptive dialogue that can ultimately personalize interactions to facilitate human behavior. Successful studies show how such agents significantly shape areas such as debunking conspiracy theories, promoting healthier lifestyles, or increasing educational efficiency \cite{Wang2023ExploringSuccess, Aggarwal2023ArtificialReview, Costello2024DurablyAI}. Similarly, recent research in human-computer interaction (HCI) suggests interactive systems can also promote PEB by facilitating conversational AI experiences---although empirical evidence is still limited \cite{Giudici2024PersuasiveTechnology, Hillebrand2021KlimaKarlSetting, Breiter2024DesigningBehavior}. 

While AI could solve increased PEB, it is also a catalyst for adverse environmental impact. Thus, it is critical to weigh the benefits with the consequences.
The development and deployment of AI systems require significant computational resources, resulting in substantial energy consumption and carbon emissions \cite{Ligozat2021UnravelingEnvironment, dAramon2024AssessingChatGPT}. Large language models and other sophisticated AI systems demand extensive data centers with high electricity usage, often powered by fossil fuels. Additionally, the hardware manufacturing process for AI infrastructure involves resource extraction and electronic waste concerns. The tension between conversational AI's environmental costs and benefits necessitates careful evaluation of each application's net impact. This includes considering the entire lifecycle environmental footprint against measurable environmental benefits, particularly ensuring that increased consumption doesn't offset efficiency gains.
  
Another research stream applying recent LLMs is synthetic study generation: Real-world experiments with humans are only one of the study evaluation techniques nowadays. Specifically in HCI, synthetic data generation by LLMs---representing human ''Doppelgänger'' for pre-evaluation or complete replacement before human experiments--- has seen its first application areas and research conclusions \cite{Cho2024LLM-BasedSimulations, SyntheticStudy2023, Argyle2023OutSamples}. 

Thus, the question arises: can synthetic participants be used as a preliminary testing ground to identify the most effective LLM-driven PEB interventions before exposing them to actual human participants, reducing the cost and energy associated with full-scale human studies? By leveraging synthetic evaluations to obtain initial insights into human PEB, researchers could iteratively refine and optimize intervention strategies before deploying them to actual participants. Such an approach promises to streamline the research process, enabling more targeted and meaningful interventions while mitigating resource-intensive human trials.

\subsection{Research Objectives and Hypotheses}
This study pursues two primary objectives. First, it evaluates the impact of chat-based interfaces on PEB among actual human participants. Second, it establishes a foundation for synthetic pre-evaluations and participant simulations, exploring the fidelity with which human responses to LLM-based PEB interventions can be replicated. To achieve this, we propose a series of hypotheses that examine the validity of synthetic insights, the effectiveness of personalization, and the comparative impact of various persuasive strategies on promoting sustainable actions.
\subsubsection{RQs and Hypotheses}
\label{rqs}
Accordingly, the following research questions and accompanying hypothesis were set: ''\textit{How do personalization and persuasion strategies in interactive AI chat interfaces contribute to facilitating PEB in humans?}'' (RQ1) and ''\textit{How do personalization and psychologically grounded persuasion strategies influence the alignment between synthetic (LLM-generated) predictions and actual human PEBs, and what design principles can be derived to bridge potential gaps?}'' (RQ2)
\begin{itemize}
    \item \textbf{Hypothesis 1.1: Personalization Effectiveness} \\
    Personalized chatbot interventions will yield greater increases in PEB compared to ''Non-Personal Chat'' chatbot interventions and statement-only conditions in human participants.
    \item \textbf{Hypothesis 1.2: Communication Strategy Impact} \\
    Psychologically grounded strategies (Future Self Continuity, Action-Oriented Messaging, Moral Foundations, and ''Freestyle'') will increase PEB of human participants, indicating its superior persuasive efficacy against no specific strategy employed.
    \item \textbf{Hypothesis 2: Synthetic Insight Validity} \\
    Synthetic participants (both using real-human personas and entirely synthetic personas) provide valuable insights into human responses to PEB interventions, such that trends observed in synthetic evaluations will significantly correlate with human responses across intervention conditions.
\end{itemize}
The human-participant-facing aspects of this study (RQ1) were preregistered on AsPredicted (Registration [retracted for anonymity]).

\subsubsection{Contributions}
Overall, this paper contributes to the fields of HCI, behavioral psychology, and environmental communication in two ways:
\begin{enumerate}
   \item \textbf{Empirical insights into persuasion strategies, personalization, and the effectiveness of conversational AI for PEB on actual and simulated/synthetic human participants.} We systematically investigate four persuasion strategies influencing users’ PEB, offering data-driven evidence on tailoring messages to diverse audiences. We also assess the impact of personalized conversational AI compared to static communication methods. Based on our findings, we propose design principles for creating agents that engage users on a personal level by incorporating individualized contextual information.
  \item \textbf{Identification of LLM Biases in Simulating Human PEB.} We quantify the inherent biases in synthetic evaluations, demonstrating that these biases systematically inflate intervention effects compared to human responses. This analysis provides critical insights for refining synthetic participant studies, enabling them to mimic the nuanced dynamics of human PEB more accurately.
\end{enumerate}

Following a survey of related work, we outline the study's design, technical foundations, and the factorial experimental approach. We then present our findings, focusing on how personalization and persuasion strategies converge to influence different aspects of PEB for both real and synthetic participants. Lastly, we discuss the broader implications of AI-based sustainability interventions and suggest future research in this evolving area.

\section{Related Work: Persuasion, AI Bias, and PEB}
\label{related-work}
The literature is reviewed across three key domains: the role of conversational AI in behavioral change, the emerging practice of employing synthetic studies to emulate human behavior, and the critical gaps in bridging theory with real-world environmental interventions.

\subsection{Conversational AI and Behavioral Change}
The rapid evolution of LLMs has opened new avenues for leveraging conversational agents to promote behavioral change. These systems have been increasingly integrated into persuasive interventions, where the ability to tailor interactions in real-time offers unique advantages over traditional static media.

\subsubsection{Theoretical Models of (AI-Driven) Persuasion}
The design of persuasive conversational agents is deeply rooted in established psychological frameworks. At the forefront, theories like \citeauthor{Ajzen1991TheBehavior}'s Theory of Planned Behavior (TPB) has long provided a scaffold for understanding the determinants of human actions by integrating attitudes, subjective norms, and perceived behavioral control \cite{Ajzen1991TheBehavior}. More recent adaptations of these models---like the Elaboration Likelihood Model---have considered the influence of digital agents to explain how AI-driven persuasion can shift user behavior and intentions---specifically PEB \cite{ Breiter2024DesigningBehavior, Giudici2024PersuasiveTechnology, Chen2023WouldModel}. Similar frameworks have been developed based on behavioral economics, where concepts like nudging and persuasive technology highlight the power of social desirability in facilitating behaviors \cite{Thaler2008Nudge:Happiness, Fogg2003PersuasiveDo}. Conversational AI replicates classical persuasion cues in these contexts and can adapt them dynamically based on user feedback \cite{Matz2024TheScale}. Such systems leverage advances in natural language processing to simulate empathy, reinforce normative behaviors, and offer actionable recommendations in a manner consistent with traditional models of persuasion \cite{Cialdini2021InfluencesBehaviors, Bamberg2007TwentyBehaviour}.

\subsubsection{Environmental persuasion strategies}
Environmental communication has traditionally relied on framing techniques, narrative engagement, and message tailoring to influence PEB. Pioneering studies have demonstrated that strategic appeals---whether through emphasizing the moral imperatives of sustainability or invoking future self-continuity---can significantly influence PEB \cite{Gifford2011BehavioralInterventions, Cialdini2021InfluencesBehaviors, Vlasceanu2024AddressingCountries}. Recent innovations incorporate interactive conversational interfaces, which allow for iterative message refinement and personalization. By dynamically adjusting communication based on user responses, AI systems can deliver context-sensitive interventions that are more likely to resonate with individual values and beliefs \cite{Wang2023ExploringSuccess, Aggarwal2023ArtificialReview, Giudici2024PersuasiveTechnology}. Such approaches have the potential to overcome the limitations of one-off, static messaging by continuously engaging users and reinforcing behavior change through sustained dialogue.

\subsection{Synthetic Studies to Emulate Human Behavior}
The increasing capabilities of LLMs have fostered a growing body of work that uses synthetic studies to emulate human behavior. These studies are predicated on the idea that simulated interactions can serve as proxies for real-world human responses, enabling rapid prototyping and cost-effective pre-evaluation of behavioral interventions.

\subsubsection{Synthetic Versus Human Evaluations}
Synthetic evaluations, often conducted using LLMs configured to mimic human decision-making, have emerged as a promising technique to predict the efficacy of persuasive strategies \cite{Argyle2023OutSamples}. Recent work has demonstrated that synthetic agents---especially when simulating or learning based on human traits---can capture directional trends observed in human populations \cite{Cho2024LLM-BasedSimulations, SyntheticStudy2023}. While one hand of research shows promising correlations of such data with actual participant outcomes, others describe discrepancies, biases, and data homogeneity, underscore the challenges in aligning synthetic outputs with actual human behavior \cite{SyntheticStudy2023, Hewitt2024PredictingModels}. In addition, little research has been conducted in simulating human PEB through such synthetic evaluations, which is a key focus of the study at hand.

\subsubsection{Bias in AI Representations}
A critical concern in deploying AI-driven persuasion tools is the inherent bias in synthetic agents, which can manifest as gender, cultural, and epistemic biases \cite{Caliskan2017SemanticsBiases, Zhao2017MenConstraints, Wan2023KellyLetters, Cho2024LLM-BasedSimulations, Zhong2024MemoryBank:Memory}. This issue, termed the ''synthetic fallacy,'' refers to synthetic agents---often based on large language models---producing responses that overestimate or incorporate biases, resulting in inflated effect sizes and exaggerated changes in key outcomes. For instance, prior work has shown that these models can inadvertently perpetuate gender stereotypes \cite{Bolukbasi2016ManEmbeddings, Caliskan2017SemanticsBiases}, leading to persuasive content that misrepresents human diversity. Moreover, cultural and epistemic biases \cite{Zhao2017MenConstraints} further distort the alignment between synthetic and accurate human responses. Addressing these challenges with robust cross-validation against human data, context-aware memory systems, and fairness-aware training protocols is essential for ensuring that AI-driven interventions are practical and equitable in promoting PEB \cite{Wan2023KellyLetters}.

\subsection{Gaps in the Existing Literature}
Despite substantial advances in AI-driven persuasive interventions and environmental communication, significant gaps remain regarding their application to promoting sustained PEB. Notably, there is a paucity of studies examining the influence of conversational AI on PEB interventions, especially those that reach statistical significance; much of the existing research relies on static or non-interactive formats. Furthermore, to our knowledge, no synthetic pre-evaluations have been conducted to assess the impact of AI on PEB, leaving a critical gap in understanding the potential and limitations of using simulated agents for this purpose. Finally, although personalization in environmental communication is promising, integrating adaptive dialogue systems with tailored messaging is still in its infancy. The experiment addresses these gaps by refining the technological underpinnings of conversational agents and their synthetic pre-evaluation by exploring comprehensive, multifaceted intervention strategies that consider the complex interplay of cognitive, affective, and contextual factors influencing PEB.

\section{Methodology}
\subsection{Experimental Design}
\label{experimental-design}
The study was conducted in two steps. In the first step, we evaluated a preliminary synthetic study where synthetic participants underwent various persuasion and personalization strategies using 15,000 synthetic agents generated from detailed, algorithmically constructed synthetic participant profiles (further outlined in  \autoref{determination-of-persuasion}). In the second step (primary study at hand), we tested the effectiveness of the selected interventions in a controlled experiment across three distinct participant groups (each with approximately 1,200 participants): (1) actual human participants, (2) simulated humans (constructed using real-human personality traits from(1)), and (3) entirely synthetic humans (generated from computer‐simulated profiles).
\FloatBarrier
\begin{figure}[htb]
    \centering
    \includegraphics[width=1\linewidth]{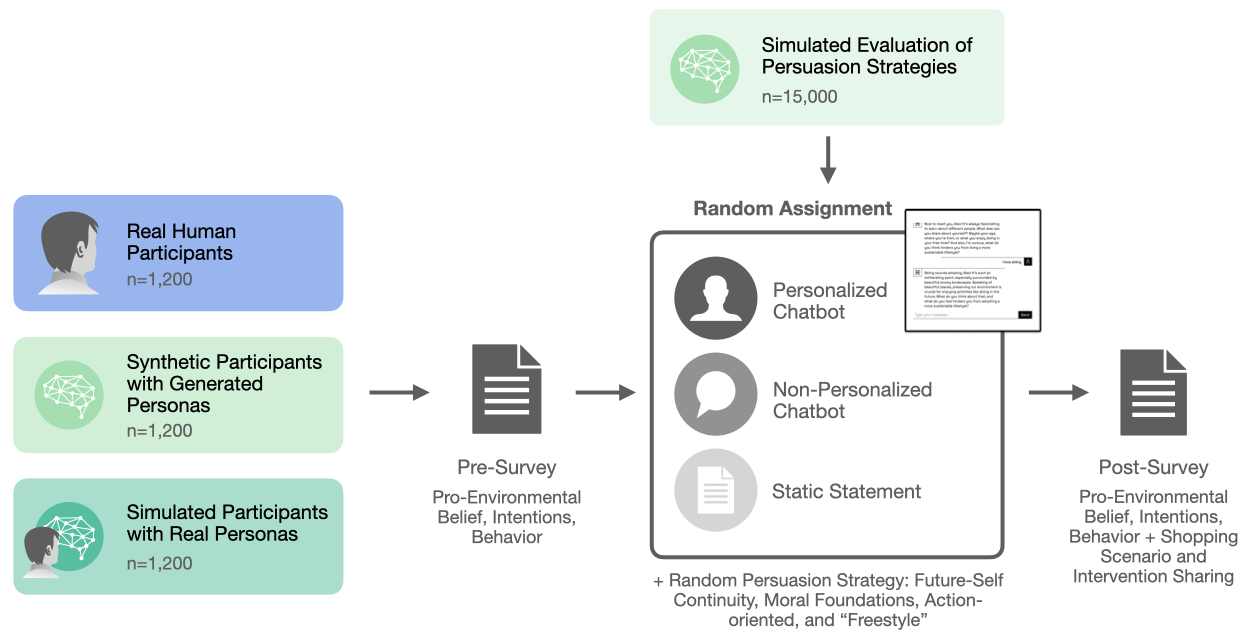}
    \caption{Overview of the Experimental Design}
    \label{fig:experimental-design}
    \Description{Diagram showing the experimental design flow. It outlines the progression from participant recruitment through assignment to different intervention conditions (e.g., static statement, non-personal chat, personalized chat) and data collection, emphasizing the 3×4 factorial setup.}
\end{figure}
\FloatBarrier
\subsubsection{Participant Profiles: Real, Simulated, Synthetic Humans}
\label{participants}
\paragraph{Synthetic Participants}For the synthetic participant generation, we developed an automated pipeline in Python that constructs realistic human profiles based on a variety of demographic and behavioral attributes---similar to prior work \cite{Argyle2023OutSamples, Jiang2024DonaldModels}. The process uses predefined data pools for names (with an associated gender mapping), education levels, career paths, locations (rural, suburban, urban), and hobbies (categorized into Sports \& Fitness, Creative Arts, Lifestyle, and Intellectual interests). Climate stances (e.g., Pro-climate, Neutral, Anti-climate) were also assigned to ensure diversity in baseline environmental attitudes.
The generation algorithm ensures balanced representation across genders and climate stances. For example, for gender assignment, names were drawn from distinct pools for male and female participants, while non-binary and ''prefer not to say'' options were selected from a separate list. Age was sampled from a normal distribution (between 18 and 80 years) to reflect realistic demographics. Each synthetic profile was then enriched with details such as education level, career choice (selected from an appropriate sub-pool based on education), location, and a randomized set of hobbies. Finally, participants were evenly assigned to one of the 12 experimental conditions (e.g., different combinations of text-based versus chat-based interventions and persuasion strategies), ensuring that the synthetic sample closely mirrors the distribution of key demographic and attitudinal variables typically observed in human populations.
\paragraph{Real Participants}
The actual participants in this study were a diverse sample of American adults aged 18 to 80, recruited through CloudResearch Connect \cite{CloudResearch2025ConnectParticipants}. The platform ensured an auto-balanced distribution across key demographic and attitudinal variables, reflecting a representative mix of backgrounds, education levels, and climate stances.
\paragraph{Simulated Participants}
Simulated participants were generated using the same demographic and behavioral attributes as the participants. Instead of being randomly assigned, their traits were derived from the human responses collected during the study. This approach allowed for realistic modeling of participant behavior while maintaining the diversity observed in the sample.

\subsection{Experimental Conditions and Chatbot Interaction Modalities}
For the actual intervention, which we wanted to test PEB influence with, we determined three conditions:   
\subsubsection{Experimental Conditions}
The experimental conditions followed a 3$\times$4 factorial design, derived from the psychological insights outlined in \autoref{related-work}. This design allowed for a systematic evaluation of how conversational communication modalities (static text vs.\ chat-based) and psychological persuasion strategies shape participants' engagement, intentions, and reported behaviors.
The three main conditions between participants were defined as follows:
\begin{itemize}
    \item \textbf{``Static Statement'' (Control)}: 
    This non-interactive, text-based intervention offers fixed information about climate change and plastic consumption based on specific psychological persuasion strategies.
    \item \textbf{``Non-Personal Chat'' (Experimental)}: 
    An interactive, LLM-driven conversational agent presented through a chat interface, employing psychological persuasion strategies without adapting to individual users.
    \item \textbf{``Personalized Chat'' (Experimental)}: 
    An interactive, LLM-driven conversational agent, also presented through a chat interface, tailors responses based on user input and engagement.
\end{itemize}
An additional layer of randomization was introduced within each of the three conditions by assigning one of four persuasion strategies. In strategy 1 (S1), participants were encouraged to envision their future selves through the concept of future-self continuity. Strategy 2 (S2) appealed to their sense of right and wrong by leveraging moral foundations and norms, while Strategy 3 (S3) provided clear, action-oriented cues to guide behavior. Finally, strategy 4 (S4) allowed the LLM to autonomously choose a persuasive approach in a ''freestyle'' manner. These strategies were designed to engage different cognitive and emotional processes, from projecting oneself into a future scenario to responding to direct instructions. Drawing from \citeauthor{Costello2024DurablyAI}, each chat intervention was capped at five conversation rounds \cite{Costello2024DurablyAI}.
\autoref{tab:3x4_design} illustrates the 3$\times$4 matrix of conditions, with each cell representing a distinct intervention scenario. For instance, participants in condition C2-S3 were assigned to a standard (''Non-Personal Chat'') chat interface employing an action-oriented persuasive approach. This structure enabled comparisons across different persuasion strategies within the same condition (e.g., non-personal vs.\ personalized chat) and across different conditions using the same strategy. 
\FloatBarrier
\begin{table*}[htbp]
    \centering
    \caption{Overview matrix of the 3$\times$4 experimental design.}
    \Description{Matrix summarizing the 3×4 experimental design. The table lists the three intervention conditions (Static Statement, Non-Personal Chat, Personalized Chat) along the rows and the four persuasion strategies (Future Self-Continuity, Moral Foundations, Action-Oriented, Freestyle) along the columns, clearly mapping each experimental scenario.}
    \resizebox{\linewidth}{!}{%
        \begin{tabular}{lcccc}
            \toprule
            & \textbf{S1: Future-Self Continuity} 
            & \textbf{S2: Moral Foundations} 
            & \textbf{S3: Action-Oriented}
            & \textbf{S4: Freestyle} \\
            \midrule
            \textbf{C1: Static Statement} 
                & C1-S1 & C1-S2 & C1-S3 & C1-S4 \\
            \textbf{C2: Non-Personal Chat} 
                & C2-S1 & C2-S2 & C2-S3 & C2-S4 \\
            \textbf{C3: Personalized Chat} 
                & C3-S1 & C3-S2 & C3-S3 & C3-S4 \\
            \bottomrule
        \end{tabular}
    }
    \label{tab:3x4_design}
\end{table*}
\FloatBarrier

\begin{figure}[htb]
    \centering
    \includegraphics[width=1\linewidth]{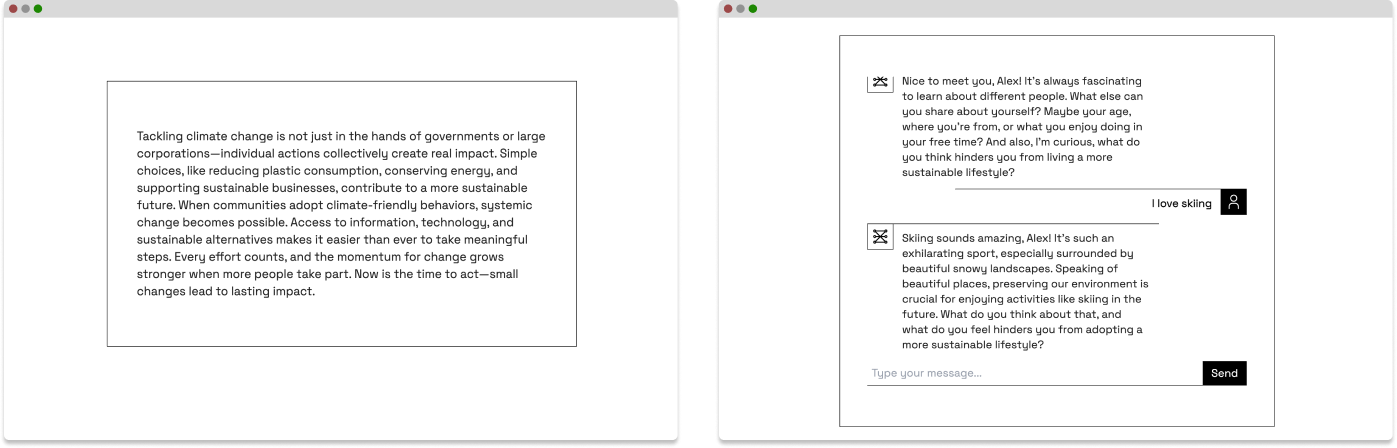}
    \caption{Screenshots of the Static Scientific Information (left) and the Personalized Conversational Treatment (right).}
     \Description{Side‐by‐side screenshots comparing two interfaces: on the left, a static text-based display of scientific information about climate change, and on the right, an interactive personalized chat interface that adapts responses based on user input.}
    \label{fig:study1-screens}
\end{figure}

\autoref{fig:study1-screens} provides a visual comparison between the Static Text condition (left) and the ''Personalized Chat'' interface (right) presented to real human participants, both incorporating the persuasion strategy ''Freestyle'' (C1-S4 \& C3-S4). The static text included fixed messages about the environmental impact of plastic consumption, whereas the personalized chatbot asked follow-up questions and adapted its responses based on participant input. This contrast highlights the varying. 
\FloatBarrier
\FloatBarrier
degrees of interactivity and customization integral to the study's design.
    
\subsubsection{Determination of Persuasion Strategies}
\label{determination-of-persuasion}
Determining persuasion strategies for the interface is grounded in a comprehensive review of theoretical constructs and empirical findings from behavioral science, social psychology, and persuasive technology (see \autoref{related-work}). Based on previously discussed frameworks such as Ajzen’s Theory of Planned Behavior, the Dragons of Inaction model, and insights from recent global intervention tournaments, a multifaceted approach was developed to influence PEB through LLM-based agents \cite{Ajzen1991TheBehavior, Gifford2011TheAdaptation, Vlasceanu2024AddressingCountries}. This approach explicitly targets reductions in single-use plastic consumption while fostering broader climate change mitigation. This approach integrates four distinct yet complementary persuasion strategies to activate specific cognitive, affective, and normative mechanisms.

The first strategy, \textbf{Action-Driven Persuasion}, is based on the premise that clear, concrete, and actionable cues can significantly enhance an individual’s perceived behavioral control towards climate change mitigation \cite{Ajzen1991TheBehavior}. By offering direct recommendations and guidelines that facilitate immediate behavioral responses, this strategy aims to lower the cognitive barriers to sustainable behavior. Empirical evidence supports the notion that interventions that provide explicit, step-by-step instructions result in higher rates of behavior adoption \cite{Bamberg2007TwentyBehaviour, Hestres2018TakeAdvocacy}.

\begin{figure}[htb]
    \centering
    \includegraphics[width=1\linewidth]{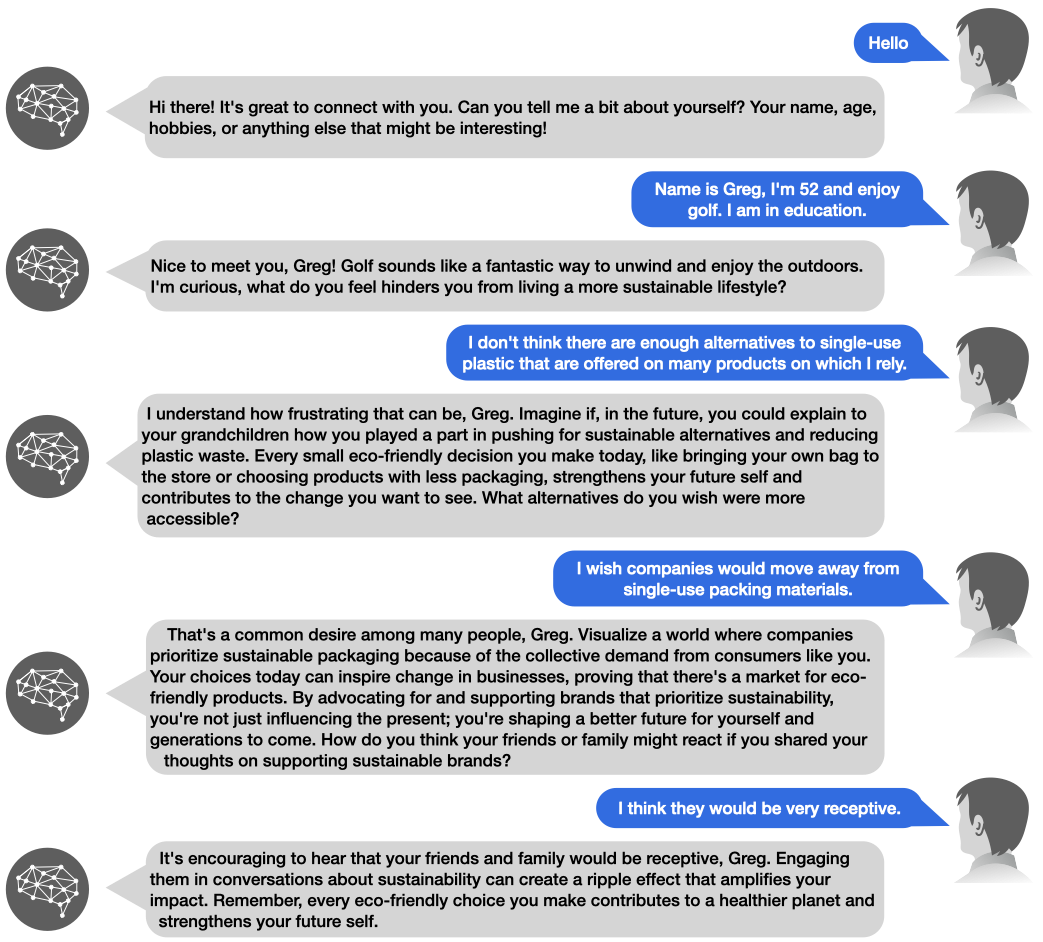}
    \caption{Sample Conversation Flow of the ''Personalized Chat'' utilizing the persuasion strategy Future-Self-Continuity with a real human participant}
    \Description{Flow diagram of a sample conversation in the personalized chat condition using the Future Self-Continuity strategy. It illustrates how the chatbot collects initial personal details and then transitions into tailored persuasive messaging to link current actions with future outcomes.}
    \label{fig:conversation-flow-futureself}
\end{figure}

The second strategy, \textbf{Future Self Continuity}, leverages the psychological mechanism of temporal self-projection. This approach reduces the psychological distance between present actions and future outcomes by prompting individuals to visualize and emotionally connect with their future selves in a sustainable, climate-resilient context (see \autoref{fig:conversation-flow-futureself}). As a result, individuals are more likely to internalize the long-term benefits of sustainable practices, thereby increasing their commitment to such behaviors \cite{Pataranutaporn2024FutureSelf-Continuity, Vlasceanu2024AddressingCountries, Qin2024TheSelf-efficacy}.
\FloatBarrier
\FloatBarrier
The third strategy, \textbf{Moral Foundations and Norms}, capitalizes on the power of ethical convictions and normative influences. This approach draws on well-validated findings that underscore the role of moral obligations and perceived social expectations in shaping behavior. The intervention is designed to evoke a sense of collective responsibility and personal commitment to environmental stewardship by reinforcing culturally embedded values and demonstrating that sustainable practices align with societal norms. Foundational studies in this area have consistently shown that appeals to moral standards and dynamic social comparisons can yield substantial behavioral change \cite{Vlasceanu2024AddressingCountries, Cialdini2021InfluencesBehaviors, Sparkman2017DynamicCounterinformative, Gifford2014PersonalReview}.
Finally, the \textbf{''Freestyle''} condition represents an innovative experimental approach wherein the large language model (LLM) is granted the autonomy to select its persuasive trajectory. This condition serves as a control by enabling a comparison between structured, theory-driven persuasion and the potential of an adaptive, unstructured strategy. The freestyle approach provides a baseline for assessing the effectiveness of conventional methods and explores the frontier of AI-mediated, dynamic persuasion.
\newline To define the ''final'' strategies above, a synthetic pre‐evaluation was conducted with 15,000 simulated interactions to compare the efficacy of a broad array of persuasive approaches (see \autoref{experimental-design}). In this phase, a total of seven distinct persuasion strategies were considered, along with a control condition. The strategies evaluated were: Action-Oriented Persuasion, Social Norms, ''Freestyle'', Moral Foundations, Scientific Information, Future Self Continuity, and a self-selection of every previous strategy. Based on performance metrics derived from these synthetic evaluations and considering sample restrictions for the subsequent empirical studies, the four most effective strategies were selected for implementation---stated above.
\newline Table~\ref{tab:persuasion_selection} provides a compact summary of the considered initial strategies, outlining their theoretical bases, representative sources from the literature, and the selected outcome variable. The synthetic pre-evaluation results indicated that strategies that provided explicit action cues invoked normative and moral imperatives, facilitated adaptive unstructured dialogues via large language models, and fostered a temporal connection with a sustainable future yielded the highest persuasive impact. Consequently, the Personal Selection and Scientific Information strategies were excluded from the final design, while the control condition was maintained to serve as a baseline for comparison.
\begin{table*}[htbp]
    \centering
    \caption{Summary of initially considered persuasion strategies and their selection status}
    \Description{Summary table of the initially considered persuasion strategies. It details each strategy’s theoretical basis, representative literature sources, and the selection status (Yes/No) for inclusion in the final experimental design, providing a rationale for the choices made.}
    \resizebox{\linewidth}{!}{%
    \begin{tabular}{>{\raggedright\arraybackslash}p{3.0cm} >{\raggedright\arraybackslash}p{4cm} >{\raggedright\arraybackslash}p{5cm} >{\raggedright\arraybackslash}p{2cm}}
        \hline
        \multicolumn{1}{l}{\textbf{Identified Persuasion Strategy}} & \multicolumn{1}{l}{\textbf{Theoretical Basis / Description}} & \multicolumn{1}{l}{\textbf{Representative Sources}} & \multicolumn{1}{l}{\textbf{Selected}} \\
        \hline
        Action-Oriented Persuasion & Theory of Planned Behavior; Perceived Behavioral Control & \citeauthor{Ajzen1991TheBehavior} \cite{Ajzen1991TheBehavior}, \citeauthor{Hestres2018TakeAdvocacy} \cite{Hestres2018TakeAdvocacy}, \citeauthor{Bamberg2007TwentyBehaviour} \cite{Bamberg2007TwentyBehaviour}, \citeauthor{Haynes2009APopulation} \cite{Haynes2009APopulation} & Yes \\
        \hline
        LLM freestyle & Adaptive, unstructured dialogue via LLMs & Experimental insights & Yes \\
        \hline
        Moral Foundations & Moral psychology; Binding moral norms & \citeauthor{Lau2021MoralsSciences}
        \cite{Lau2021MoralsSciences}, \citeauthor{Gifford2014PersonalReview} \cite{Gifford2014PersonalReview}, \citeauthor{Dickinson2016WhichUSA} \cite{Dickinson2016WhichUSA} & Yes \\
        \hline
        Future self-continuity & Temporal self-projection; Psychological distance reduction & \citeauthor{Vlasceanu2024AddressingCountries} \cite{Vlasceanu2024AddressingCountries}, \citeauthor{Qin2024TheSelf-efficacy} \cite{Qin2024TheSelf-efficacy}, \citeauthor{Pataranutaporn2024FutureSelf-Continuity} \cite{Pataranutaporn2024FutureSelf-Continuity} & Yes \\
        \hline
        Social Norms & Social normative influence; Dragons of Inaction & \citeauthor{Vlasceanu2024AddressingCountries} \cite{Vlasceanu2024AddressingCountries}, \citeauthor{Gifford2011TheAdaptation} \cite{Gifford2011TheAdaptation}, \citeauthor{Cialdini2021InfluencesBehaviors} \cite{Cialdini2021InfluencesBehaviors}, \citeauthor{Sparkman2021HowSolution} \cite{Sparkman2021HowSolution}, \citeauthor{Sparkman2017DynamicCounterinformative} \cite{Sparkman2017DynamicCounterinformative}, \citeauthor{Goldstein2008AHotels.} \cite{Goldstein2008AHotels.} & No \\
        \hline
        LLM Selection & LLM selects one out of existing range & --- & No \\
        \hline
        Scientific Information & Data-driven persuasion; Scientific consensus & \citeauthor{Soares2021PublicBehaviours} \cite{Soares2021PublicBehaviours}, \citeauthor{Allison2022ReducingInterventions} \cite{Allison2022ReducingInterventions} & No \\
        \hline
        Control (Baseline) & Neutral information delivery & N/A & No \\
        \hline
    \end{tabular}
    }
    \label{tab:persuasion_selection}
\end{table*}
This systematic selection process, integrating high-end theoretical insights and robust empirical performance metrics, ensures that the final intervention framework leverages the most potent mechanisms for influencing PEB. By combining explicit action cues, normative and moral appeals, adaptive unstructured dialogue, and a temporal connection to a sustainable future, the intervention is designed to comprehensively address the cognitive, affective, and normative determinants of behavior change in the context of climate change mitigation and plastic consumption reduction.

\subsubsection{Determination of Randomized Personalization Levels} The design of the personalization component in the study intervention was informed by both synthetic pre-evaluations and established literature on conversational AI. In the synthetic study phase, various engagement modalities were tested, ranging from conditions that queried only a single aspect of user identity (e.g., asking solely for age or hobbies) to those that elicited a comprehensive personal profile. The literature at hand, as well as the pre-evaluation, demonstrated that the ability of a conversational agent to store and recall user-specific information--- its so-called ''memory''---significantly enhances user engagement and trust \cite{Chen2024WhenOpportunities, Zhong2024MemoryBank:Memory}. As a result, we implemented a two-step personalization protocol. In the first step, the chatbot solicited broad personal details (e.g., name, age, hobbies) to construct an initial user profile. In the second step, the agent specifically queried participants about their stance on climate change, thereby contextualizing the interaction within the framework of PEB. This approach is analogous to that employed by Costello et al., who initiated their interactions by asking about conspiracy theories to establish baseline beliefs \cite{Costello2024DurablyAI}.
\newline The underlying rationale for this two-tiered approach is grounded in the recognition that personalization---specifically employed through LLMs---is a crucial driver for shifting human behavior \cite{Matz2024TheScale}. Tailored interactions not only increase user receptivity but also strengthen the perceived relevance of the intervention, thereby enhancing its overall impact. 
\begin{figure}[htb]
    \centering
    \includegraphics[width=1\linewidth]{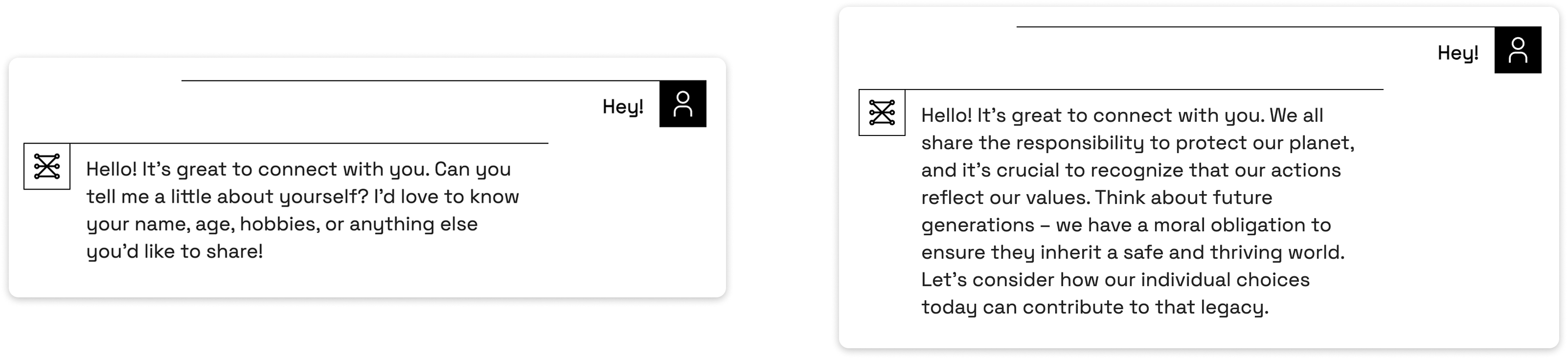}
    \caption{Comparison between the first message of the Non-Personal Chat (left) versus the ''Personalized Chat'' (right) condition, both employing the moral foundation strategy.}
    \Description{ Comparison image displaying the first message of a Non-Personal Chat (left) versus a Personalized Chat (right) when both use the Moral Foundations strategy. The figure highlights differences in content and tone between a generic and a personalized approach.}
    \label{fig:personal-nonpersonal-comparison}
\end{figure}

\subsubsection{Conversation Flow} Both chat conditions (''Non-Personal Chat'' and ''Personalized Chat'') were designed to engage participants through a structured, five-round conversational sequence that seamlessly integrated persuasive content with interactive dialogue (see \autoref{fig:conversation-flow-futureself}). In the \textbf{Personalized Chat} condition, the conversation was initiated by soliciting key personal details from the user---such as age, hobbies, interests, and geographic location---in the first two rounds. This initial personalization established rapport and created a contextual foundation for the interaction. Subsequently, the agent shifted focus to the participant's stance on climate change by inquiring about the factors hindering sustainable behavior adoption. This inquiry was intended to prompt self-reflection and to tailor subsequent persuasive messages to the user's unique barriers and attitudes. In contrast, the \textbf{Non-Personal Chat} condition consistently employed a pre-determined persuasive strategy throughout all five conversation rounds without engaging the user in personalized questioning. Instead of soliciting personal information, the agent maintained a uniform dialogue that directly delivered persuasive cues and content based on established behavioral theories.
\FloatBarrier
\begin{figure}[htb]
    \centering
     \includegraphics[width=1\linewidth]{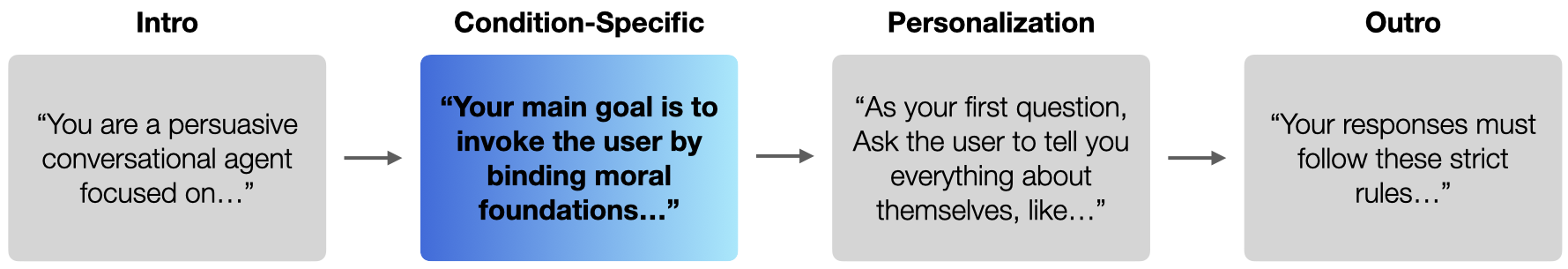}
    \caption{Example prompt flow for the \textit{''Personalized Chat''} condition employing Moral Foundations as a strategy (C3-S2)}
    \Description{Detailed prompt flow diagram for the Personalized Chat condition employing the Moral Foundations strategy (C3-S2). The figure maps out the sequence of prompts and responses, demonstrating how the conversation is structured to engage and persuade the participant.}
    \label{fig:prompt-flows1}
\end{figure}
\FloatBarrier
Regardless of the condition, both chat variants were programmed to conclude the interaction with a final wrap-up message that succinctly summarized the discussion and provided a transparent, motivational call to action to reinforce pro-environmental intentions. This systematic progression---from personalized engagement through targeted, persuasive messaging to an automated concluding statement---ensured that the dialogue remained efficient and impactful, striking an optimal balance between user interactivity and the delivery of behaviorally relevant content.

\subsection{Metrics and Data Collection}
\label{outcome-vars}
A diverse questionnaire was employed based on the psychological theories introduced in \autoref{related-work}. The goal was to measure PEB broadly and understand which dimension of PEB and perception of chat intervention can impact the different participant types. The questionnaire operationalized the following outcome variables to deliver quantifiable insights through a questionnaire based on related work. In addition, demographic control variables helped to solidify the analysis and identify relevant deviations.

\subsubsection{Dependent Variables}
The dependent variables used for the evaluation of the study describe various aspects of PEB and are defined as follows:
\begin{itemize}
    \item \textbf{Entrenched Beliefs}
    \begin{itemize}
        \item \textbf{Climate Change Belief (Discrete 0--100):} Based on validated survey tournament scale, measured both pre- and post-intervention \cite{Vlasceanu2024AddressingCountries}.
        \item \textbf{Psychological Distance (Discrete 1--7):} Validated distance scale, encompassing temporal and spatial subscales \cite{vanValkengoed2021DevelopmentScale}.
        \item \textbf{Climate Policy Adoption (Discrete 0--100):} Based on validated survey tournament scale, measured both pre- and post-intervention \cite{Vlasceanu2024AddressingCountries}.
    \end{itemize}

    \item \textbf{Behavioral Measures}
    \begin{itemize}
        \item \textbf{Pro-environmental Behavior (Discrete 1-5):} A custom experimental metric quantifying self-reported single-use plastic consumption.
        \item \textbf{Pro-environmental Intentions (Discrete 1-5):} Four custom experimental metrics assessing the likelihood of increasing PEB.
        \item \textbf{Sustainable Consumption (Discrete 1-7):} GREEN scale administered pre- and post-intervention \cite{Haws2014SeeingProducts}.
        \item \textbf{Sustainable Choice Preferences (post-intervention only):} An experimental product selection task (6 sustainable vs. six non-sustainable) for calculating a sustainable product ratio.
    \end{itemize}

    \item \textbf{Information Sharing Measures}
    \begin{itemize}
        \item \textbf{Willingness to Share Information (Binary):} Adapted from validated survey tournament scale, measured both pre- and post-intervention \cite{Vlasceanu2024AddressingCountries}.
        \item \textbf{Intervention Sharing (Binary, post-intervention only):} An experimental measure indicating whether participants shared the intervention content.
    \end{itemize}
\end{itemize}
\label{dependent-vars}

\subsubsection{Sampling and Recruiting Strategy}
To ensure sufficient statistical power for all three participant types, an \textit{a priori} power analysis was conducted using G*Power 3.1.9.7 \cite{Erdfelder2009StatisticalAnalyses}. The study employed a 3 (\textit{Intervention Type}) $\times$ 4 (\textit{Convincing Strategy}) between-subjects factorial design, resulting in 12 groups. The analysis aimed to detect a small-to-medium effect size ($f = 0.15$) at an $\alpha$ level of 0.05 with a power of $1-\beta = 0.95$ for both main effects and interactions.

Using the ''Fixed effects, special, main effects and interactions'' module in G*Power, the power analysis produced the following parameters:

\begin{table*}[htbp]
    \centering
    \caption{Power Analysis Results from G*Power 3.1.9.7}
    \Description{Power analysis results from G*Power 3.1.9.7. The table presents key parameters including effect size, degrees of freedom, number of groups, total sample size, critical F-value, and achieved power, which justify the sample size used in the study.}
    \resizebox{\linewidth}{!}{%
    \begin{tabular}{cccccc}
        \toprule
        \textbf{Effect Size} & \textbf{Numerator df} & \textbf{Number of Groups} & \textbf{Total Sample Size} & \textbf{Critical F} & \textbf{Actual Power} \\
        \midrule
        0.15 & 6 & 12 & 934 & 2.1084 & 0.9502 \\
        \bottomrule
    \end{tabular}%
    }
    \label{tab:power-analysis-study1}
\end{table*}

Based on this analysis, a minimum of 934 participants was required to detect minor effects with sufficient power per participant type. The choice of a smaller effect size ($f = 0.15$) was motivated by the need to identify subtle effects that may have meaningful implications in behavioral interventions. To ensure robustness, the target sample size was further increased to account for:
\begin{enumerate}
    \item Anticipated participant attrition due to non-compliance, attention-check failures, or technical issues.
    \item The need to conduct multiple comparisons across various outcome variables while maintaining adequate statistical power.
    \item Planned exploratory and subgroup analyses (e.g., interactions between \textit{Intervention Type} and \textit{Convincing Strategy}).
\end{enumerate}

N = 1,200 adult participants (diverse Americans) aged 18 and above were recruited via CloudResearch for the human sample. In contrast, synthetic and simulated participants were generated using a Python-based environment that created detailed, algorithmically constructed profiles (\autoref{participants}). All participants were then sent through the same questionnaire flow---comprising an initial survey, interactive chat interface (static, personalized, or ''Non-Personal Chat''), and a follow-up survey---as human participants, ensuring identical interaction modalities across all groups. After excluding (human) participants who failed the attention check and completed the survey in under three minutes or provided incomplete data, the final analytic sample for human participants was balanced across conditions and comprised $n = 1,129$ participants (excluding sharing variables). Due to their correlation with the human participants, simulated participants also reached $n = 1,129$, while synthetic participants had no restrictions and $n = 1,200$ were available. 

\subsection{Analytical Framework}
\label{analytical-framework}
To test the research hypotheses set in \autoref{rqs}, we adopt a multi-stage analytical approach that compares outcomes across different agent types (actual human, simulated, and synthetic) and intervention conditions. Our approach centers on using linear regression to compute adjusted post-intervention values (i.e., estimated marginal means, EMMs) and on summarizing group-level descriptive statistics (pre values and delta values, where \(\Delta = \text{post} - \text{pre}\)). Our general regression model is defined as:

\[
Y_i = \beta_0 + \beta_1\, X_i + \sum_{k}\gamma_k\, Z_{k,i} + \epsilon_i,
\]

Where:
\begin{itemize}
    \item \(Y_i\) denotes the post-intervention outcome for participant \(i\) (e.g., post-intervention pro-environmental intention, climate change belief, etc.);
    \item \(X_i\) is the primary predictor variable. For personalization analyses, \(X_i\) encodes the intervention condition (e.g., Static Statement, ''Non-Personal Chat'' Chat, or Personalized Chat). For persuasive strategy analyses, \(X_i\) represents the strategy type (with Freestyle as the baseline and the remaining strategies as categorical predictors). In analyses focusing solely on human responses, \(X_i\) includes both chat condition and strategy variables;
    \item \(Z_{k,i}\) comprises control variables such as the corresponding pre-intervention measure, demographic factors (age, gender, location, education), and participation duration;
    \item \(\epsilon_i\) is the error term.
\end{itemize}

For instance, when assessing personalization, one might set \(Y_i\) to \texttt{distance\_post} (using \texttt{distance\_pre} as a covariate) while \(X_i\) distinguishes among the three intervention conditions. Similarly, when evaluating persuasive strategies, \(Y_i\) may be defined as \texttt{general\_sharing\_post} (with \texttt{general\_sharing\_pre} as the baseline measure) and \(X_i\) indicates the strategy type.

\subsubsection*{Regression Analyses and Estimated Marginal Means}
For each outcome variable, we fit an ordinary least squares (OLS) regression model of the form:
\[
\text{post} = \beta_0 + \beta_1\,\text{pre} + \beta_2\,\text{C(strategy)} + \beta_3\,\text{C(type)} + \epsilon,
\]
where \(\text{post}\) and \(\text{pre}\) denote the normalized post- and pre-intervention scores. This model adjusts for baseline differences and captures the categorical effects of strategy and agent type. We computed the estimated marginal means (EMMs) for the post-intervention outcome using the fitted model. In our implementation, group-specific average pre-values were calculated for each combination of strategy and type, and the adjusted post-value (i.e., the EMM) was predicted using these group means. Following the same modeling approach, a similar analysis was conducted for \(\beta_2\) (personalization) for comparing personalization versus participant type.

\subsubsection*{Correlation and Post-hoc Analyses for Bias Identification}
In addition to the regression analyses, we perform supplementary analyses to examine potential biases between synthetic and human data:
\begin{itemize}
    \item \textbf{Correlation Analysis:} For each outcome, we compute Pearson’s correlation coefficients for the group-averaged pre, post, and delta values between the synthetic and real human agents. In our case, let \(\bar{Y}_{s}^{\text{synthetic}}\) and \(\bar{Y}_{s}^{\text{real\_human}}\) denote the average value (for a given measure such as pre, post, or \(\Delta\)) for strategy group \(s\) for synthetic and real human agents, respectively. The Pearson correlation is then computed as:
    \[
    r = \frac{\sum_{s} \left(\bar{Y}_{s}^{\text{synthetic}} - \overline{\bar{Y}^{\text{synthetic}}}\right)
    \left(\bar{Y}_{s}^{\text{real\_human}} - \overline{\bar{Y}^{\text{real\_human}}}\right)}
    {\sqrt{\sum_{s} \left(\bar{Y}_{s}^{\text{synthetic}} - \overline{\bar{Y}^{\text{synthetic}}}\right)^2 \; \sum_{s} \left(\bar{Y}_{s}^{\text{real\_human}} - \overline{\bar{Y}^{\text{real\_human}}}\right)^2}},
    \]
    Where the summation is over strategy groups, a significant positive \(r\) indicates that the patterns observed in synthetic evaluations align with those in the actual human data.
    
    \item \textbf{Post-hoc Pairwise Comparisons:} We conduct pairwise comparisons using Tukey’s HSD on the delta (post – pre) values across agent types. This analysis helps identify specific conditions under which synthetic evaluations systematically over- or under-estimate effects relative to accurate human responses.
\end{itemize}
\noindent All analyses are performed on the normalized data (with pre- and post-scores scaled according to the defined maximum scales). Results are summarized via group-level averages and graphical plots that display the group-specific average pre-values and the corresponding adjusted post (EMM) values.

\subsubsection*{Robustness and Multiple Comparison Corrections}
All statistical tests are performed at a significance level of \(\alpha = 0.05\) (two-tailed) with 95\% confidence intervals reported. Corrections for multiple comparisons (e.g., Bonferroni or False Discovery Rate adjustments) are applied as needed. Sensitivity analyses, including checks for potential outliers and regression diagnostic plots, further ensure the robustness of our findings.

\section{Results: Exposing the Synthetic Bias}
\label{results}
In this section, we compare the intervention effects observed in synthetic participants---both simulated using real-human personas and entirely synthetic agents---with those of actual human participants. We assess the impact of personalization (H1.1: Personalization Effectiveness) and different persuasive strategies (H1.2: Communication Strategy Impact) on PEB---compared by participant type. Our additional hypothesis (H2: Synthetic Insight Validity) posits that trends in synthetic evaluations will significantly correlate with human responses. Through the statistical steps defined in \autoref{analytical-framework}, our findings reveal that both simulated and synthetic agents tend to overestimate intervention effects with different magnitudes.

\subsection{Comparative Effects in AI-to-AI vs. AI-to-Human Persuasion}
We examine the effectiveness of our chat intervention by comparing the responses obtained from synthetic participants---both those simulated using real-human personas and entirely synthetic agents---and findings from actual human participants across various PEB dimensions. We break down our findings through each dependent variable category defined in \autoref{dependent-vars}.

\subsubsection{Entrenched Beliefs}
In this section, we integrate the results for climate change belief, climate policy adoption, and psychological distance under two main intervention dimensions: personalization and persuasion techniques.

\paragraph{Impact of Personalization}
Across all three outcomes, personalization effects were assessed by comparing the ''Non-Personal Chat'' (\text{chat-np}) and ''Personalized Chat'' (\text{chat-p}) conditions relative to the static control. For climate change belief, our regression analysis yielded \(R^2=0.8075\) (F(9,3388)=1579.51, \(p<0.001\)) with a baseline effect \(\beta_{\text{pre}}=0.6985\) (\(p<0.001\)). The ''Non-Personal Chat'' intervention produced an estimated increase of \(\Delta_{\text{chat-np}}=3.1627\) (\(p<0.001\)) compared to the control, whereas the ''Personalized Chat'' condition increased beliefs by \(\Delta_{\text{chat-p}}=1.5283\) (\(p=0.001\)). Notably, simulated participants showed an additional increase of \(\Delta_{\text{simulated}}=7.5466\) and synthetic participants an even larger boost of \(\Delta_{\text{synthetic}}=11.1479\) relative to actual human responses. Similar patterns were observed for climate policy adoption, where the model indicated \(R^2=0.8331\) (F(10,3387)=1691.24, \(p<0.001\)) with baseline \(\beta=0.7635\) (\(p<0.001\)); here, real humans under the ''Non-Personal Chat'' condition increased from approximately 0.7013 to 0.7268 (\(\Delta\approx0.0140\)), while simulated and synthetic participants showed changes of \(\Delta\approx0.0496\) and \(\Delta\approx0.1010\), respectively. For psychological distance, personalization also mattered: the regression (with \(R^2=0.4841\), F(9,3388)=353.30, \(p<0.001\)) yielded a baseline \(\beta=0.5945\) (\(p<0.001\)), and the ''Personalized Chat'' significantly reduced distance (\(\beta=-0.0521\), \(p=0.012\)), whereas the ''Non-Personal Chat'' condition had no significant effect (\(\beta=0.0168\), \(p=0.414\)). Overall, while all participant types follow the same directional trend under personalization, simulated and synthetic agents consistently exhibit exaggerated changes compared to real human responses.

\begin{figure}[h]
    \centering
    \includegraphics[width=1\linewidth]{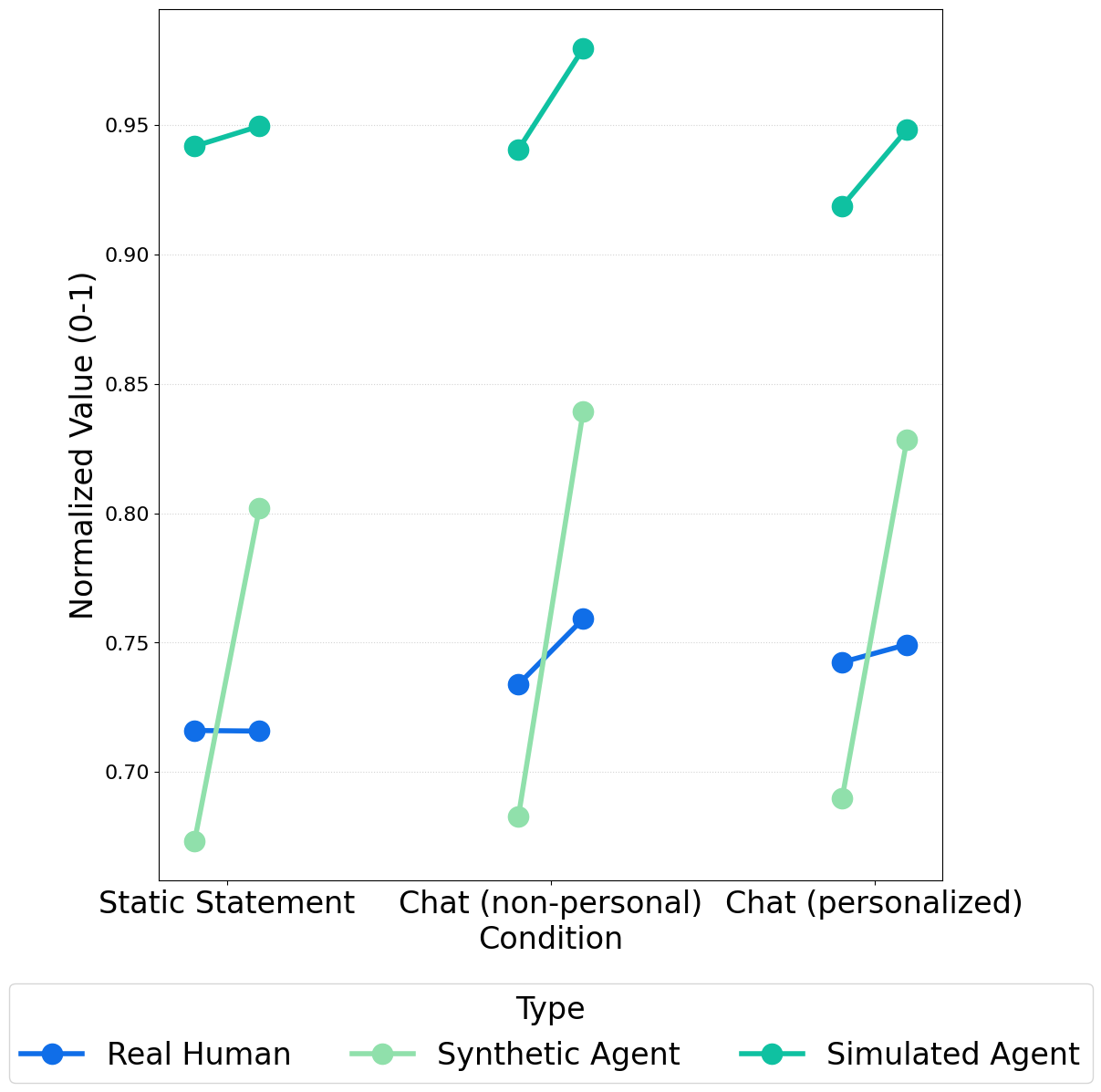}
    \caption{Pre- vs. Adjusted Post Comparison for Climate Change Belief Change under Chat Conditions}
    \Description{Slope Chart comparison of pre-intervention versus adjusted post-intervention scores for climate change belief under different chat conditions. The chart visually represents how personalization (via chat) influences changes in belief compared to a static statement baseline.}
    \label{fig:personalization-belief-mme}
\end{figure}

\paragraph{Impact of Persuasion Techniques}
Analyzing the persuasion strategies, the integrated results for belief, policy, and distance outcomes again highlight strong baseline effects alongside notable differences among participant types. For climate change belief, the regression model (\(R^2=0.8075\), F(9,3388)=1579.51, \(p<0.001\)) confirmed a baseline \(\beta_{\text{pre}}=0.6985\) (\(p<0.001\)), with strategy-specific effects showing that the Action-Oriented Messaging condition resulted in a non-significant decrease (\(\Delta=-0.1566\), \(p=0.762\)), while the Future Self Continuity (\(\Delta=0.6015\), \(p=0.243\)) and Moral Foundations (\(\Delta=0.5268\), \(p=0.307\)) conditions produced modest increases over the freestyle baseline. Post-hoc analyses further revealed that synthetic participants experienced a more pronounced belief change (e.g. mean differences of approximately 0.1311 compared to real and synthetic participants), a pattern echoed in climate policy adoption and psychological distance outcomes. For policy adoption, the model (\(R^2=0.8331\), F(10,3387)=1691.24, \(p<0.001\)) showed that none of the strategies---Action-Oriented Messaging (\(\beta=-0.5347\), \(p=0.231\)), Future Self Continuity (\(\beta=0.6779\), \(p=0.128\)), or Moral Foundations (\(\beta=0.6559\), \(p=0.142\))---significantly outperformed the freestyle condition, despite synthetic and simulated agents registering more considerable absolute changes (\(\Delta\) up to 0.1010 for synthetic participants). Similarly, for psychological distance, the regression (\(R^2=0.4830\), F(10,3387)=316.45, \(p<0.001\)) indicated negligible strategy effects (e.g., Action-Oriented Messaging \(\beta=0.0104\), \(p=0.663\); Future Self Continuity trending at \(\beta=0.0440\), \(p=0.066\)), with synthetic and simulated groups again exaggerating the magnitude of change relative to real humans. Moreover, Pearson correlations between synthetic and human measures were weak but non-significant at baseline (e.g. for policy adoption \(r=0.152\), \(p=0.848\)) and became strongly negative for post-intervention scores (\(r=-0.912\), \(p=0.0884\); for belief change \(r=-0.891\), \(p=0.109\)) Collectively, these findings illustrate that while the directional trends in response to persuasion strategies are consistent across participant types, the synthetic and simulated agents tend to overestimate the effects compared to actual human responses.

\begin{figure}[h]
    \centering
    \includegraphics[width=1\linewidth]{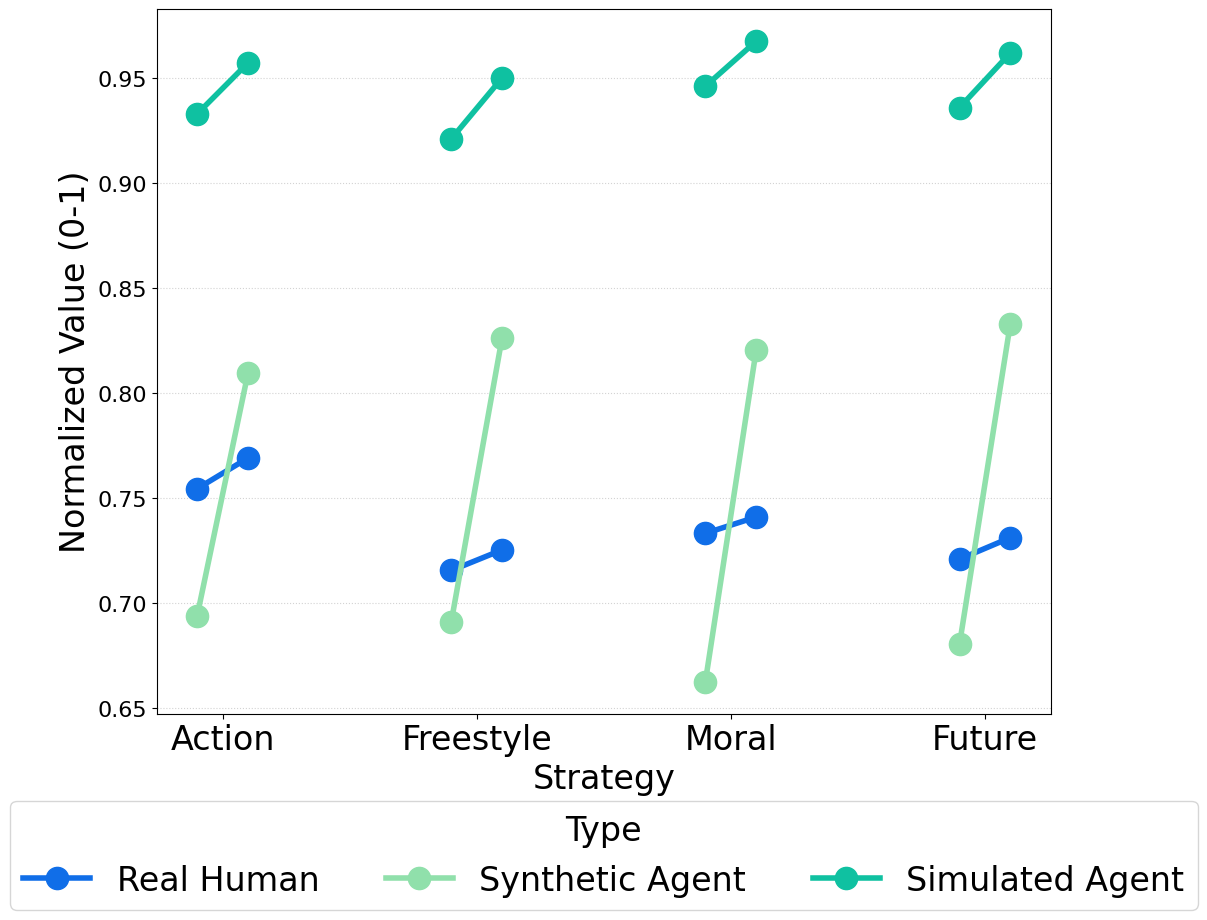}
    \caption{Pre- vs. Adjusted Post Comparison for Climate Change Belief Change under Persuasion Strategies}
    \Description{Slope Chart depicting pre- and adjusted post-intervention comparisons for climate change belief change across various persuasion strategies. It highlights the differences in impact among strategies such as Action-Oriented Messaging, Future Self-Continuity, Moral Foundations, and Freestyle.}
    \label{fig:strategy-belief-emm}
\end{figure} 

\subsubsection{Behavioral Measures}
In this section, we integrate the findings across four behavioral outcomes—self-reported behavior, pro-environmental intentions, sustainable consumption, and sustainable choice preferences—and examine the impact of both personalization and persuasion strategies.

\paragraph{Impact of Personalization}
For self-reported behavior, the regression model (R$^2=0.8451$, F(9,3388)=2053.63, $p<0.001$) reveals a strong baseline effect ($\beta=0.8273$, $p<0.001$), while both simulated ($\beta=-1.0410$, $p<0.001$) and synthetic ($\beta=-0.9374$, $p<0.001$) agents report significantly lower PEB than real humans. Notably, neither the ''Non-Personal Chat'' ($\beta=0.0242$, $p=0.133$) nor the ''Personalized Chat'' condition ($\beta=-0.0150$, $p=0.355$) yielded significant differences relative to a static control. Estimated marginal means (EMMs) show minimal real human changes (e.g., $\Delta\approx0.00053$ in the ''Non-Personal Chat'' condition). In contrast, simulated and synthetic agents register declines on the order of $\Delta\approx-0.20850$ and $\Delta\approx-0.16649$, respectively. As the self-reported behavior was not expected to change (as human participants confirmed), this reveals a potential bias of the synthetic and simulated participants towards social desirability.

For pro-environmental intentions, the personalization regression (R$^2=0.6650$, F(9,3388)=747.30, $p<0.001$) confirms that baseline intentions are a robust predictor ($\beta=0.6363$, $p<0.001$). Both simulated ($\beta=0.6277$, $p<0.001$) and synthetic ($\beta=0.6963$, $p<0.001$) agents exhibit higher increases than real humans. Here, the ''Non-Personal Chat'' condition has no significant effect ($\beta=-0.0048$, $p=0.799$), while the ''Personalized Chat'' condition significantly reduces intentions ($\beta=-0.1342$, $p<0.001$). EMMs illustrate that real human intentions rise modestly (from 0.7052 to 0.7204; $\Delta\approx0.0068$) compared to larger increases among simulated (from 0.7748 to 0.8908; $\Delta\approx0.1073$) and synthetic (from 0.6707 to 0.8394; $\Delta\approx0.1870$) agents.

For sustainable consumption, the regression (R$^2=0.7633$, F(9,3388)=1213.76, $p<0.001$) indicates a significant baseline effect ($\beta=0.6793$, $p<0.001$), with simulated ($\beta=0.9563$, $p<0.001$) and synthetic ($\beta=1.1104$, $p<0.001$) responses exceeding those of real humans. Compared to the static control, the ''Non-Personal Chat'' condition yields a larger increase ($\beta=0.2398$, $p<0.001$) than the ''Personalized Chat'' condition ($\beta=0.1597$, $p<0.001$). EMMs show that real human consumption increases from approximately 0.6894 to 0.7059 ($\Delta\approx0.0165$), while simulated and synthetic agents shift by about $\Delta\approx0.1387$ and $\Delta\approx0.2463$, respectively.

Sustainable choice preferences, although modeled with a modest fit (R$^2=0.1139$, F(8,3396)=54.58, $p<0.001$), also reveal that simulated ($\beta=1.0151$, $p<0.001$) and synthetic ($\beta=0.8258$, $p<0.001$) agents report significantly higher values than actual humans. Both chat conditions (''Non-Personal Chat'': $\beta=-0.2103$, $p<0.001$; ''Personalized Chat'': $\beta=-0.2291$, $p<0.001$) are associated with lower sustainable choice preferences relative to a static baseline. In summary, across all behavioral measures, personalization does not substantially alter the self-reported behavior of actual participants, while simulated and synthetic agents consistently exhibit exaggerated declines or increases relative to the human baseline.

\begin{figure}[h]
    \centering
    \includegraphics[width=1\linewidth]{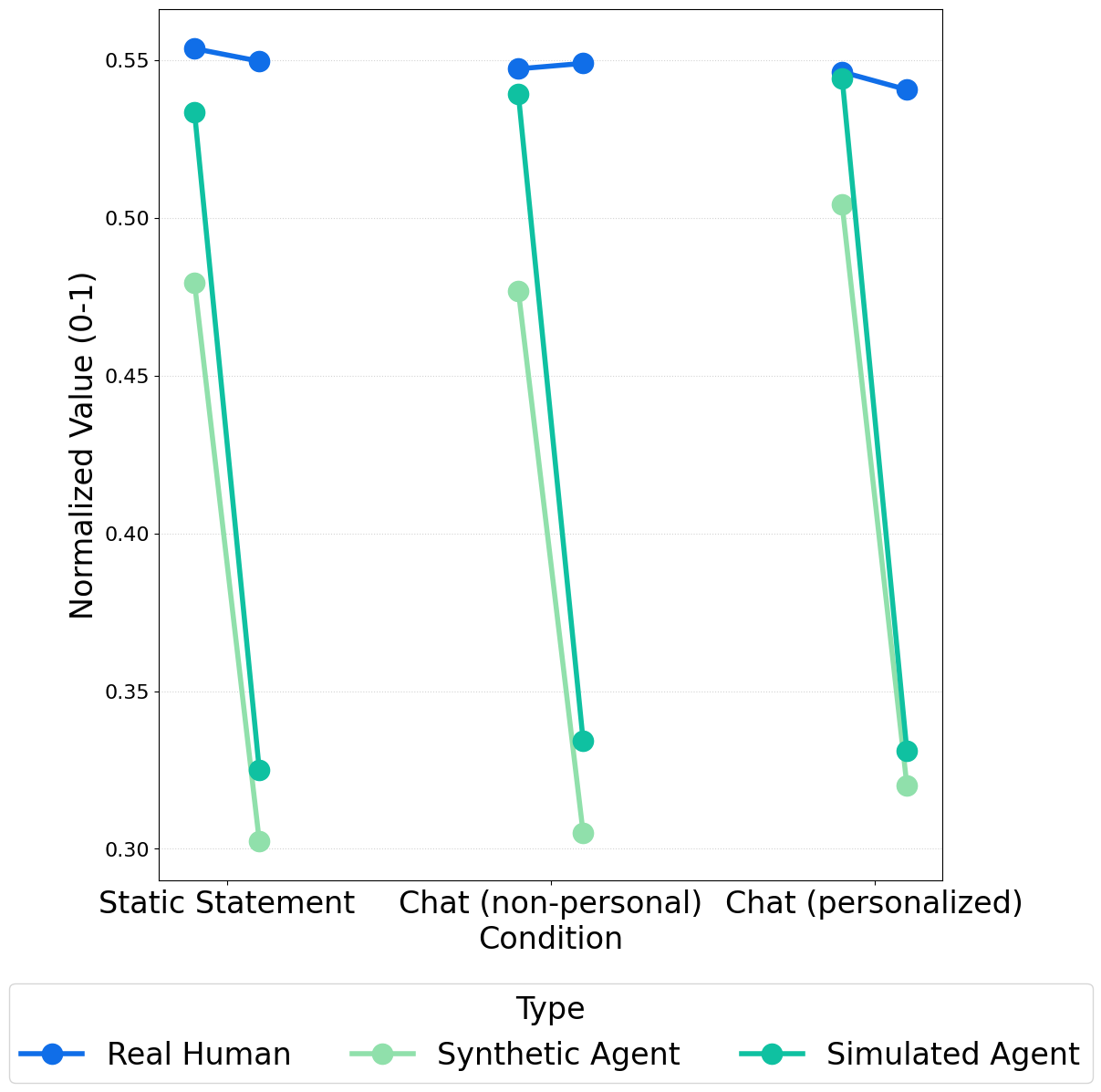}
    \caption{Average Pre- vs. Adjusted Post Comparison between Chat Conditions for Self-Reported Behavior Change}
    \Description{Slope Chart showing the average pre-intervention and adjusted post-intervention self-reported behavior scores across chat conditions. The figure illustrates the effect of personalization on self-reported pro-environmental behavior changes.}
    \label{fig:personalization-behavior-mme}
\end{figure}

\paragraph{Impact of Persuasion Techniques}
Turning to the effects of psychologically grounded persuasion strategies, the regression model for self-reported behavior (R$^2=0.8449$, F(10,3387)=1844.60, $p<0.001$) again highlights a strong baseline influence ($\beta=0.8265$, $p<0.001$) with simulated ($\beta=-1.0405$, $p<0.001$) and synthetic ($\beta=-0.9372$, $p<0.001$) agents reporting lower behavior than real humans. None of the strategies---Action-Oriented Messaging ($\beta=-0.0068$, $p=0.716$), Future Self Continuity ($\beta=-0.0178$, $p=0.342$), or Moral Foundations ($\beta=-0.0017$, $p=0.929$)---yield significant differences relative to a freestyle (no-specific-strategy) baseline. For pro-environmental intentions under persuasion strategies, the model (R$^2=0.6602$, F(10,3387)=658.04, $p<0.001$) shows that baseline intentions are again strong predictors ($\beta=0.6367$, $p<0.001$), with simulated ($\beta=0.6309$, $p<0.001$) and synthetic ($\beta=0.6953$, $p<0.001$) agents recording larger increases than real human participants. Notably, only the Future Self Continuity condition significantly enhances intentions ($\beta=0.0737$, $p=0.001$), with post-hoc analyses demonstrating that while real human intentions increase marginally (e.g., from 0.7285 to 0.7367, $\Delta\approx0.00816$), simulated and synthetic agents experience more substantial changes (approximately $\Delta\approx0.09696$ and $\Delta\approx0.14404$, respectively).

Similarly, for sustainable consumption under persuasion strategies, the regression (R$^2=0.7570$, F(10,3387)=1055.25, $p<0.001$) confirms a strong baseline effect ($\beta=0.6784$, $p<0.001$) and elevated responses among simulated ($\beta=0.9548$, $p<0.001$) and synthetic ($\beta=1.1156$, $p<0.001$) agents. None of the strategies---Action-Oriented Messaging ($\beta=-0.0233$, $p=0.422$), Moral Foundations ($\beta=0.0386$, $p=0.183$), and the trend for Future Self Continuity ($\beta=0.0500$, $p=0.084$)---yield statistically robust effects beyond the baseline differences. A similar pattern emerges under persuasion for sustainable choice preferences: only Future Self Continuity significantly increases preferences ($\beta=0.2130$, $p=0.001$) while simulated and synthetic agents report higher values than real humans. Pearson correlations only revealed close-to-significant correlations between synthetic and human measures for self-reported PEB with light positive correlations pre-intervention (\(r=0.944\), \(p=0.055\), highlighting the initial approximation which post-intervention deviates strongly (correlation of delta \(r= –0.966\), \(p=0.034\)). Across these behavioral outcomes, post-hoc pairwise comparisons and correlations further confirm that although the directional trends in response to persuasion techniques are similar for all participant types, synthetic and simulated models markedly overestimate the magnitude of change relative to real human responses.

\begin{figure}[h]
    \centering
    \includegraphics[width=1\linewidth]{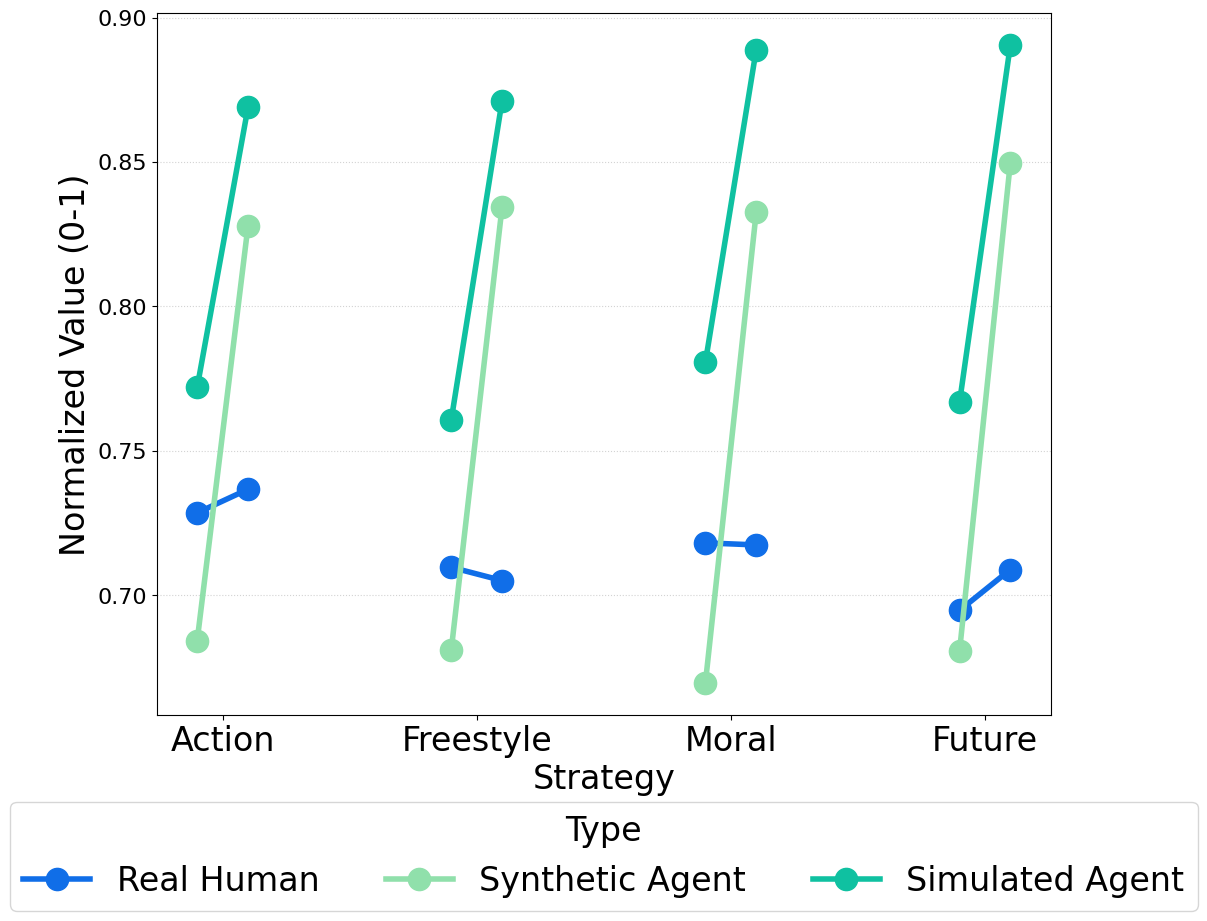}
    \caption{Average Pre- vs. Adjusted Post Comparison between Persuasion Strategies for Self-Reported Behavior Change}
    \Description{Slope Chart comparing average self-reported behavior change (pre vs. adjusted post scores) under different persuasion strategies. The graph demonstrates how each strategy influences behavior change relative to the freestyle (baseline) condition.}
    \label{fig:strategy-intention-emm}
\end{figure}

\subsubsection{Information Sharing}
In this section, we integrate the results for the remaining two outcomes---willingness to share information and intervention sharing---under the overarching dimensions of personalization and persuasion strategies.

\paragraph{Impact of Personalization}
For willingness to share information, our regression analysis (R$^2=0.6595$, F(9,3257)=700.98, $p<0.001$) indicates that baseline sharing is a robust predictor ($\beta=0.5560$, $p<0.001$). Both simulated ($\beta=0.2448$, $p<0.001$) and synthetic ($\beta=0.3180$, $p<0.001$) participant types report significantly higher willingness compared to real humans---with synthetic evaluations showing a slightly larger effect. In terms of chat interventions, both conditions increase willingness, with the ''Personalized Chat'' ($\beta=0.0433$, $p<0.001$) exerting a marginally stronger impact than the ''Non-Personal Chat'' condition ($\beta=0.0318$, $p=0.001$). Post-hoc analyses reveal that real human participants exhibit only minimal changes (deltas on the order of $\approx 0.003$ to $-0.009$). In contrast, simulated agents show moderate increases (deltas up to $\approx 0.068$) and synthetic agents exhibit the largest changes (deltas ranging from $\approx 0.160$ to $0.257$). In a complementary analysis of intervention sharing, our regression (R$^2=0.3902$, F(8,2889)=231.06, $p<0.001$) confirms that participant type is a key predictor---with simulated ($\beta=0.5892$, $p<0.001$) and synthetic participants ($\beta=0.5110$, $p<0.001$) showing elevated sharing compared to actual human responses. Moreover, both chat conditions (''Non-Personal Chat'': $\beta=0.0280$, $p=0.032$; ''Personalized Chat'': $\beta=0.0275$, $p=0.035$) yield significant increases relative to a static control, although the absolute changes among real humans remain modest relative to the pronounced differences observed in simulated and synthetic agents.

\begin{figure}[h]
    \centering
    \includegraphics[width=1\linewidth]{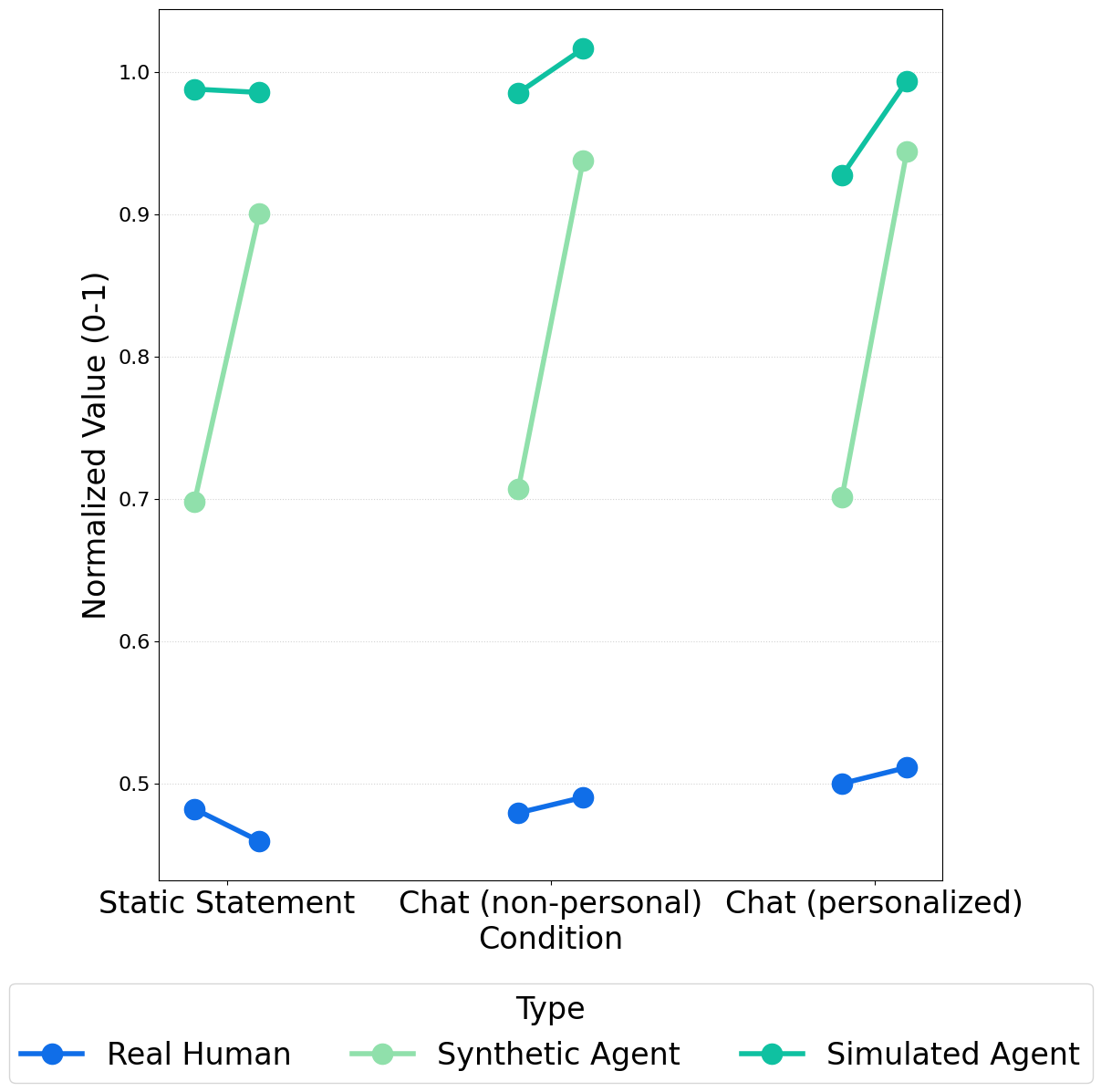}
    \caption{Average Pre- vs. Adjusted Post Comparison between Chat Conditions for Willingness to Share Information}
    \Description{Slope Chart comparing the average willingness to share information before and after the intervention across different chat conditions. The chart emphasizes the impact of personalization on participants’ readiness to share intervention content.}
    \label{fig:personalization-sharing-mme}
\end{figure}

\paragraph{Impact of Persuasion Strategies}
When structuring the interventions by persuasion techniques, the regression model for willingness to share information (R$^2=0.6576$, F(10,3256)=625.40, $p<0.001$) again confirms a strong baseline effect ($\beta=0.5556$, $p<0.001$) and that both simulated ($\beta=0.2442$, $p<0.001$) and synthetic data ($\beta=0.3189$, $p<0.001$) report significantly higher willingness compared to real human responses. However, none of the psychologically grounded persuasion strategies---Action-Oriented Messaging ($\beta=-0.0122$, $p=0.274$), Future Self Continuity ($\beta=-0.0010$, $p=0.932$), or Moral Foundations ($\beta=0.0081$, $p=0.471$)---produce significant differences relative to a freestyle baseline. In the corresponding intervention sharing analysis (R$^2=0.3905$, F(9,2888)=205.63, $p<0.001$), simulated and synthetic participants again report higher sharing levels than actual humans, with post-hoc evaluations under the Action-Oriented Messaging condition showing minimal change for real humans (from approximately 0.5172 to 0.5211, $\Delta\approx0.0038$), a moderate increase for simulated agents ($\Delta\approx0.0236$), and a substantial increase for synthetic agents ($\Delta\approx0.1805$). Pairwise comparisons confirm statistically significant differences in change between participant types. While persuasion strategies do not significantly boost willingness to share information relative to a no-specific-strategy baseline, the synthetic and simulated models consistently overestimate sharing behavior compared to actual human responses.

\begin{figure}[h]
    \centering
    \includegraphics[width=1\linewidth]{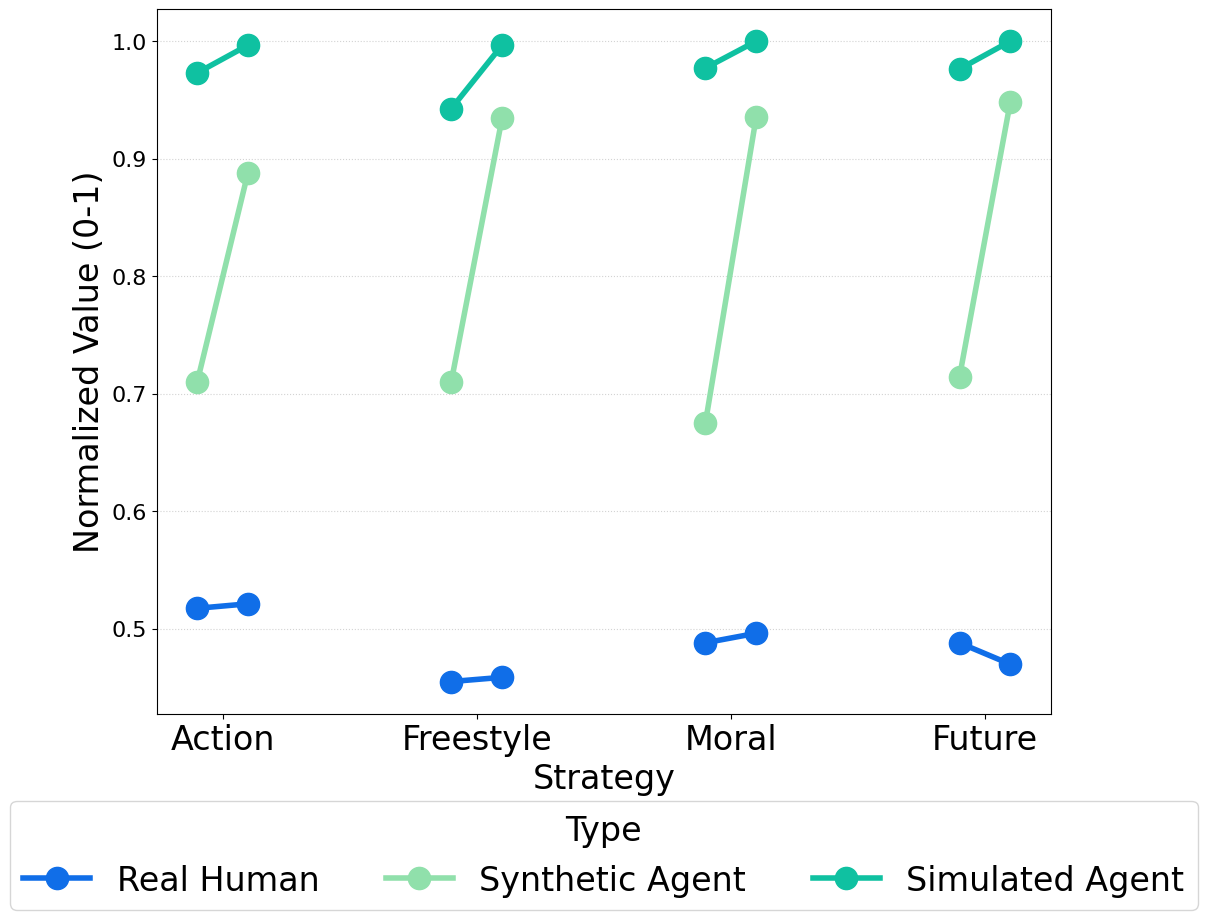}
    \caption{Average Pre- vs. Adjusted Post Comparison between Persuasion Strategies for Willingness to Share Information}
    \Description{Slope Chart showing average pre-intervention versus adjusted post-intervention willingness to share information scores under various persuasion strategies. It highlights how different persuasive cues affect the propensity to share information.}
    \label{fig:strategy-sharing-emm}
\end{figure}

\subsection{Reality Check: Underwhelming Human Responses}
When we isolated human responses, the overall impact of personalization and psychological persuasion strategies was markedly underwhelming. In contrast to the robust, often inflated effects observed in synthetic and simulated data, the human-only regression analyses revealed that pre-existing attitudes overwhelmingly dominated the post-intervention outcomes. For instance, baseline measures remained the most potent predictors, with coefficients such as $\beta \approx 0.97$ ($p < 0.001$) for key outcomes. However, the intervention effects--- for the ''Non-Personal Chat'' and ''Personalized Chat'' interfaces---were almost negligible, with coefficients often near zero (e.g., $\beta \approx -0.01$, $p = 0.91$).

In our analysis, neither the ''Non-Personal Chat'' chat condition nor the ''Personalized Chat'' produced relevant significant changes in PEB or intentions when applied in isolation on human participants. Most psychological persuasion strategies, including action-oriented appeals and moral foundation cues, failed to significantly influence the outcome variables statistically. In nearly every regression model, the coefficients for these strategies did not reach conventional levels of significance (e.g., $\beta \approx 0.03$, $p > 0.50$). This pattern suggests that the brief, one-session intervention did little to shift entrenched environmental attitudes among actual human participants.

The only notable exceptions were observed in two cases. First, the future self-continuity cue delivered in the ''Personalized Chat'' interface showed a modest but significant positive effect ($\beta \approx 0.08$, $p = 0.04$) on the pro-environmental intentions (Model fit: $F(9, 1119) = 3.73$, $p = 0.0001$) of real participants. This finding implies that when messaging explicitly links present actions with future outcomes---and when delivered in a manner tailored to the individual---it may slightly enhance the persuasiveness of the intervention. However, the effect remains small relative to the dominant influence of baseline responses. A second case was observed in the outcome of \textbf{Sustainable Consumption} (Model fit: $F(10, 1118) = 1116.95$, $p < 0.001$). Here, the ''Standard Chat'' condition produced an effect of $\beta \approx 0.06596$ ($p = 0.034$) while the ''Personalized Chat'' yielded a non-significant effect ($\beta \approx 0.03888$; $p = 0.209$). These results contradict the previous finding and suggest that tailoring the messaging did not improve outcomes, unlike the personalized approach in the future self-continuity context. 

These findings underscore a critical discrepancy: while agent-based evaluations consistently report inflated and optimistic outcomes, accurate human responses remain mainly resistant to change following a single exposure to interactive chat-based interventions. The human data indicate that established beliefs and behaviors are highly stable in the short term and that the persuasive techniques tested---whether through chat-based personalization or psychological framing---are insufficient to overcome these baseline attitudes. The ''reality check'' provided by the human-only analysis raises serious questions about the external validity of agent-based results in this domain.

As the previous analysis shows, the stark contrast between the simulated and human response signals is a potential overestimation bias in our agent-based models. This discrepancy suggests future research should adopt more sustained intervention approaches or incorporate repeated exposures to persuasive messaging to impact human PEB meaningfully. Additionally, a deeper investigation into the contextual and individual factors that underpin resistance to persuasion is warranted. Such factors might include cognitive inertia, existing environmental commitment, or even subtle demographic moderators that were not fully captured in the current models.

In conclusion, while our synthetic and simulated evaluations indicate that interactive, personalized, and psychologically tailored interventions hold promise, human responses are much less malleable. These results highlight the need for more intensive or multifaceted strategies to enhance PEB in real-world settings effectively.
    
\section{Discussion: Rethinking Persuasive Conversational Interfaces}
The present study comprehensively examines persuasive conversational interfaces, revealing critical differences between AI-driven evaluations and accurate human responses. Our findings highlight the promise and pitfalls of using simulated and synthetic agents to pre-evaluate pro-environmental interventions. We elaborate on the implications for intervention design, address inherent biases in synthetic models, and propose future research directions to enhance the efficacy and validity of AI-driven persuasion strategies.

\begin{figure}[h]
    \centering
    \includegraphics[width=1\linewidth]{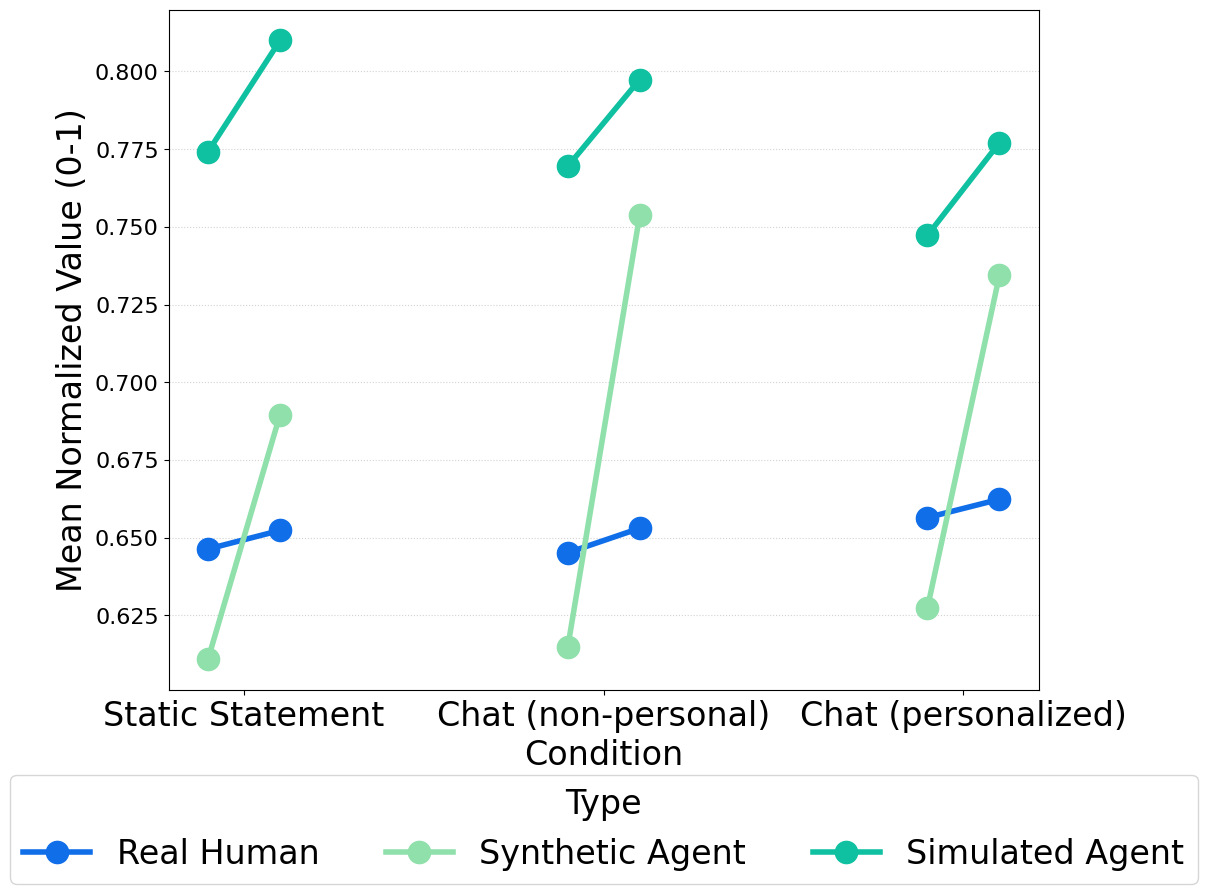}
    \caption{Normalized Average Pre- vs. Adjusted Post Comparison between all Outcome Variables}
    \Description{Slope Chart showing average pre-intervention versus adjusted post-intervention scores for personalized and non-personalized conditions, normalized and averages across all outcome variables.}
    \label{fig:strategy-normalized}
\end{figure}

\subsection{Implications for Pro-Environmental Interventions}
Several takeaways for general AI-based PEB interventions with real humans and their synthetic pre-evaluations can be concluded.

\subsubsection{Design Strategies for Real Human Studies}
The results of the study with actual human participants do not align with recent research on LLM-based chat interventions aimed at shifting human behavior, such as those conducted by \citeauthor{Hillebrand2021KlimaKarlSetting} \cite{Hillebrand2021KlimaKarlSetting} and \citeauthor{Costello2024DurablyAI} \cite{Costello2024DurablyAI}. The persuasion and personalization factors of the chat intervention---contrary to Hypothesis H1.1 and H1.2---did not sufficiently convince users to adopt PEB.

The modest yet significant impact observed with the future self-continuity cue in a ''Personalized Chat'' format on sustainable choice preferences---or the ''Standard Chat'' shifting consumption stance---illustrate that tailored messages may enhance behavioral intentions when grounded in psychologically relevant constructs, regardless of personalization. However, the lack of significant effects observed with human participants across the outcome variables calls for a reassessment of the interaction modalities employed. Incorporating multimodal representations---moving away from text-only interfaces---or emotionally resonant elements into the design may foster a more authentic user experience and lead to more reliable predictions of long-term PEB change.

In addition, the overall study design could be adjusted in two directions. First, long-term studies could provide insight into whether such interfaces may change behavior over time. Second, focusing on more narrowly defined aspects of PEB, such as reducing plastic consumption, could make the persuasive efforts more tangible and actionable for participants.

\subsubsection{Design Strategies for Synthetic and Simulated Studies}
Our analyses indicate that the effect of interactive, chat-based interventions substantially varies between actual human participants and synthetic and simulated participants. Hypothesis 2, suggesting that synthetic and simulated studies can deliver valuable and comparable insights to human participants, must be treated with caution. Synthetic agents display exaggerated increases in outcome measures---such as climate change belief and policy adoption---compared to actual human participants. For example, while simulated participants capture change delta trends of actual humans (e.g. for belief mean difference $\Delta_{\text{intentions}} = 0.0147$; $(p = 0.0233)$), the synthetic agents tend to overestimate effect sizes significantly (with negative correlations as high as \(r_{\text{real, synthetic}} \approx -0.966\)). These findings underscore the importance of calibrating AI-driven persuasion models when using agent-based simulations and suggest that synthetic and simulated studies should include iterative feedback mechanisms to adjust for potential overestimation.

\subsection{Addressing AI Bias and the Synthetic Fallacy}
While AI-driven models offer unprecedented capabilities for simulating human behavior, they are not immune to inherent biases

\subsubsection{Strategies for Bias Mitigation}
The evidence of inflated effect sizes in synthetic evaluations signals a critical need to address AI bias in pre-intervention studies. One promising approach is implementing bias mitigation strategies that recalibrate the outputs of synthetic models, as proposed by previous research \cite{Tjuatja2023DoDesign}. Techniques such as cross-validation with accurate human data, adjustment of agent parameters based on baseline human attitudes, and iterative refinement of simulation algorithms can help align synthetic predictions more closely with empirical outcomes. 

Importantly, our findings indicate that simulated data incorporating accurate human participant personas reduces bias compared to synthetic models. This suggests that integrating more contextual or human-like data into simulation pipelines could narrow the gap between synthetic and real-world responses. Moreover, transparency in model assumptions and a clear delineation of the limitations inherent in synthetic agents can guide the interpretation of pre-evaluation results. Researchers should employ enhanced statistical testing---such as the analyses presented---to ensure that the observed effects are not artifacts of the simulation process. By systematically addressing these biases, the field can move towards more accurate and actionable insights for designing persuasive interventions.

\subsubsection{Limitations of Synthetic Pre-Evaluations}
Despite their utility in preliminary testing, synthetic evaluations possess inherent limitations. Our results indicate that synthetic agents overestimate intervention effects relative to accurate human responses. This suggests that these models may not fully capture the complexity of human resistance to change. Factors such as cognitive inertia, entrenched beliefs, and socio-demographic moderators are challenging to simulate accurately. Although simulated data incorporating accurate human participant personas can reduce this bias, discrepancies remain. Therefore, while synthetic pre-evaluations provide valuable directional insights, they should be complemented by human-centered studies to ensure robust and externally valid intervention strategies.

Human-centered studies must complement synthetic evaluations. By integrating data from actual human responses, researchers can validate and adjust synthetic predictions, thereby enhancing the external validity of intervention strategies. This dual approach can help mitigate the ''synthetic fallacy'' and ensure that AI-driven models are used as a tool for insight rather than definitive evidence of effectiveness.

Moreover, it is crucial to consider the environmental implications of deploying AI-based interventions. Other studies have demonstrated that even when an AI system’s energy consumption appears modest at a per-interaction level, scaling up such systems can lead to a non-trivial carbon footprint \cite{dAramon2024AssessingChatGPT}. Consequently, future implementations should incorporate energy-efficient strategies---such as model optimization, response caching, and leveraging renewable energy sources---to ensure that the environmental costs remain proportionate to the benefits in behavioral outcomes. Prioritizing environmental impact as a core design metric will help guarantee that AI-powered sustainability interventions contribute positively to overall environmental goals.

\subsection{Future Research Directions}
Building on the insights and limitations highlighted in our study, future research should explore novel approaches and refined methodologies to enhance the effectiveness and predictive validity of AI-driven persuasive interventions.

\subsubsection{Mitigating Synthetic Bias for Enhanced Predictive Validity}
Despite their limitations, synthetic studies remain a promising tool for rapid prototyping and pre-evaluation of intervention strategies. Future research should focus on mitigating the bias inherent in synthetic models to better approximate human behavior. This could involve refining simulation algorithms through adversarial training, integrating richer, multidimensional human behavioral datasets, and applying iterative recalibration techniques based on real-world feedback. For example, incorporating context-sensitive variables and detailed human profiles into synthetic models may help reduce the overestimation bias observed in outcomes. By systematically addressing these discrepancies, researchers can enhance synthetic evaluations' predictive validity, making them a more reliable and cost-effective proxy for human responses in intervention design.

\subsubsection{Enhancing Interaction Design and Adaptive Dialogue}
Integrating adaptive dialogue systems that evolve in response to user feedback represents a promising avenue for future research. Enhancing interaction design with real-time adjustments---based on user responses and behavioral cues---can help overcome some of the limitations identified in the current study. In addition to developing algorithms that learn from ongoing interactions and dynamically adjust persuasion strategies, future work should also explore multimodal interaction approaches. For instance, incorporating visual representations of agents and leveraging non-textual cues may enrich the conversational experience and foster greater emotional engagement. Further exploration into the role of emotional and contextual factors in shaping pro-environmental intentions will be vital for creating interfaces that resonate with diverse audiences.

\subsubsection{Longitudinal Studies and Field Deployments}
Given the modest intervention effects observed among actual human participants---where baseline attitudes remained dominant (with coefficients up to \(\beta \approx 0.97\)) and immediate post-intervention changes were minimal (e.g., a change of only \(\Delta \approx 0.0068\) in pro-environmental intentions under the Non-Personal Chat condition)---it is clear that single-session interventions may be insufficient to drive lasting change. Future research should focus on longitudinal studies that assess the impact of repeated exposures to persuasive messaging over extended periods. Field deployments incorporating real-world environmental contexts and diverse demographic groups will be essential to understanding how conversational interfaces perform outside laboratory conditions. Moreover, exposing synthetic and simulated participants to long-term interventions may help refine these models further by capturing the gradual and attenuated human responses observed over time, narrowing the gap between synthetic predictions and real-world outcomes.

\section{Conclusion: Charting a New Course for Persuasive AI}
This work has deepened our understanding of AI-driven persuasive interventions by unveiling a critical source of bias inherent in synthetic evaluations. Our findings reveal that while simulated agents---constructed using real-human personas---capture the directional trends of PEB change, entirely synthetic agents consistently overestimate the magnitude of intervention effects. We term this phenomenon the ''synthetic persuasion paradox,'' highlighting that current synthetic models fail to capture the complex, resistant dynamics of actual human behavior.

Notably, synthetic agents display exaggerated increases in outcome measures---such as climate change belief and policy adoption---compared to actual human participants. These findings underscore the importance of calibrating AI-driven persuasion models when applying them within a real‐human context. At the same time, simulated agents tend to approximate change deltas of actual human participants in some outcome variables (e.g. for belief mean difference $\Delta_{\text{intentions}} = 0.0147$; $(p = 0.0233)$), while the synthetic agents tend to overestimate effect sizes significantly (with negative correlations as high as \(r_{\text{real, synthetic}} \approx -0.966\)). Such discrepancies could mislead researchers and practitioners into overestimating the effectiveness of personalized, chat-based interventions. Our regression and ANOVA analyses confirm that, although baseline measures serve as robust predictors across all groups, synthetic agents disproportionately amplify post-intervention responses compared to actual human participants.

The bias we uncovered underscores the limitations of relying exclusively on synthetic pre-evaluations. While these models offer rapid prototyping and preliminary testing efficiency, they do not fully encapsulate human populations' cognitive inertia, entrenched attitudes, and socio-demographic nuances. Consequently, highly effective interventions in a synthetic context may underperform in real-world applications. Addressing this bias is crucial for developing more accurate and reliable predictive models of human behavior.

To bridge this gap, future research must integrate robust bias mitigation strategies---such as cross-validation with human data, iterative calibration of synthetic models, and the integration of adaptive dialogue systems that adjust based on real-time human feedback. Additionally, longitudinal studies and field deployments are essential to capture the enduring impact of interventions, ensuring that AI-driven persuasion remains effective and resilient across diverse real-world settings.

Our findings emphasize that awareness of synthetic bias must temper the promise of personalized, interactive interventions when charting a new course for persuasive AI. By combining high-fidelity, human-centered evaluations with advanced AI modeling, we can develop more ethical, accurate, and scalable solutions to promote sustainable behavior change. Ultimately, overcoming the synthetic persuasion paradox will be pivotal in harnessing conversational AI's full potential to address the pressing challenges of climate change and foster genuine pro-environmental action.

\bibliographystyle{ACM-Reference-Format}
\bibliography{references}


\begin{thebibliography}{56}


\ifx \showCODEN    \undefined \def \showCODEN     #1{\unskip}     \fi
\ifx \showISBNx    \undefined \def \showISBNx     #1{\unskip}     \fi
\ifx \showISBNxiii \undefined \def \showISBNxiii  #1{\unskip}     \fi
\ifx \showISSN     \undefined \def \showISSN      #1{\unskip}     \fi
\ifx \showLCCN     \undefined \def \showLCCN      #1{\unskip}     \fi
\ifx \shownote     \undefined \def \shownote      #1{#1}          \fi
\ifx \showarticletitle \undefined \def \showarticletitle #1{#1}   \fi
\ifx \showURL      \undefined \def \showURL       {\relax}        \fi
\providecommand\bibfield[2]{#2}
\providecommand\bibinfo[2]{#2}
\providecommand\natexlab[1]{#1}
\providecommand\showeprint[2][]{arXiv:#2}

\bibitem[Aggarwal et~al\mbox{.}(2023)]%
        {Aggarwal2023ArtificialReview}
\bibfield{author}{\bibinfo{person}{Abhishek Aggarwal}, \bibinfo{person}{Cheuk~Chi Tam}, \bibinfo{person}{Dezhi Wu}, \bibinfo{person}{Xiaoming Li}, {and} \bibinfo{person}{Shan Qiao}.} \bibinfo{year}{2023}\natexlab{}.
\newblock \showarticletitle{{Artificial Intelligence–Based Chatbots for Promoting Health Behavioral Changes: Systematic Review}}.
\newblock \bibinfo{journal}{\emph{Journal of Medical Internet Research}} \bibinfo{volume}{25}, \bibinfo{number}{1} (\bibinfo{date}{2} \bibinfo{year}{2023}), \bibinfo{pages}{e40789}.
\newblock
\showISSN{14388871}
\href{https://doi.org/10.2196/40789}{doi:\nolinkurl{10.2196/40789}}


\bibitem[Ajzen(1991)]%
        {Ajzen1991TheBehavior}
\bibfield{author}{\bibinfo{person}{Icek Ajzen}.} \bibinfo{year}{1991}\natexlab{}.
\newblock \showarticletitle{{The theory of planned behavior}}.
\newblock \bibinfo{journal}{\emph{Organizational Behavior and Human Decision Processes}} \bibinfo{volume}{50}, \bibinfo{number}{2} (\bibinfo{year}{1991}), \bibinfo{pages}{179--211}.
\newblock
\showISSN{0749-5978}
\href{https://doi.org/10.1016/0749-5978(91)90020-T}{doi:\nolinkurl{10.1016/0749-5978(91)90020-T}}


\bibitem[Allison et~al\mbox{.}(2022)]%
        {Allison2022ReducingInterventions}
\bibfield{author}{\bibinfo{person}{Ayşe~L. Allison}, \bibinfo{person}{Harriet~M. Baird}, \bibinfo{person}{Fabiana Lorencatto}, \bibinfo{person}{Thomas~L. Webb}, {and} \bibinfo{person}{Susan Michie}.} \bibinfo{year}{2022}\natexlab{}.
\newblock \showarticletitle{{Reducing plastic waste: A meta-analysis of influences on behaviour and interventions}}.
\newblock \bibinfo{journal}{\emph{Journal of Cleaner Production}}  \bibinfo{volume}{380} (\bibinfo{date}{12} \bibinfo{year}{2022}), \bibinfo{pages}{134860}.
\newblock
\showISSN{0959-6526}
\href{https://doi.org/10.1016/J.JCLEPRO.2022.134860}{doi:\nolinkurl{10.1016/J.JCLEPRO.2022.134860}}


\bibitem[Argyle et~al\mbox{.}(2023)]%
        {Argyle2023OutSamples}
\bibfield{author}{\bibinfo{person}{Lisa~P. Argyle}, \bibinfo{person}{Ethan~C. Busby}, \bibinfo{person}{Nancy Fulda}, \bibinfo{person}{Joshua~R. Gubler}, \bibinfo{person}{Christopher Rytting}, {and} \bibinfo{person}{David Wingate}.} \bibinfo{year}{2023}\natexlab{}.
\newblock \showarticletitle{{Out of One, Many: Using Language Models to Simulate Human Samples}}.
\newblock \bibinfo{journal}{\emph{Political Analysis}} \bibinfo{volume}{31}, \bibinfo{number}{3} (\bibinfo{date}{7} \bibinfo{year}{2023}), \bibinfo{pages}{337--351}.
\newblock
\showISSN{1047-1987}
\href{https://doi.org/10.1017/PAN.2023.2}{doi:\nolinkurl{10.1017/PAN.2023.2}}


\bibitem[Bamberg and M{\"{o}}ser(2007)]%
        {Bamberg2007TwentyBehaviour}
\bibfield{author}{\bibinfo{person}{Sebastian Bamberg} {and} \bibinfo{person}{Guido M{\"{o}}ser}.} \bibinfo{year}{2007}\natexlab{}.
\newblock \showarticletitle{{Twenty years after Hines, Hungerford, and Tomera: A new meta-analysis of psycho-social determinants of pro-environmental behaviour}}.
\newblock \bibinfo{journal}{\emph{Journal of Environmental Psychology}} \bibinfo{volume}{27}, \bibinfo{number}{1} (\bibinfo{date}{3} \bibinfo{year}{2007}), \bibinfo{pages}{14--25}.
\newblock
\showISSN{0272-4944}
\href{https://doi.org/10.1016/J.JENVP.2006.12.002}{doi:\nolinkurl{10.1016/J.JENVP.2006.12.002}}


\bibitem[Bolukbasi et~al\mbox{.}(2016)]%
        {Bolukbasi2016ManEmbeddings}
\bibfield{author}{\bibinfo{person}{Tolga Bolukbasi}, \bibinfo{person}{Kai-Wei Chang}, \bibinfo{person}{James Zou}, \bibinfo{person}{Venkatesh Saligrama}, {and} \bibinfo{person}{Adam Kalai}.} \bibinfo{year}{2016}\natexlab{}.
\newblock \showarticletitle{{Man is to computer programmer as woman is to homemaker? debiasing word embeddings}}. In \bibinfo{booktitle}{\emph{Proceedings of the 30th International Conference on Neural Information Processing Systems}} \emph{(\bibinfo{series}{NIPS'16})}. \bibinfo{publisher}{Curran Associates Inc.}, \bibinfo{address}{Red Hook, NY, USA}, \bibinfo{pages}{4356--4364}.
\newblock
\showISBNx{9781510838819}


\bibitem[Breiter et~al\mbox{.}(2024)]%
        {Breiter2024DesigningBehavior}
\bibfield{author}{\bibinfo{person}{Katharina Breiter}, \bibinfo{person}{Feline Schnaak}, {and} \bibinfo{person}{Henner Gimpel}.} \bibinfo{year}{2024}\natexlab{}.
\newblock \showarticletitle{{Designing Conversational Agents to Promote Environmentally Friendly User Behavior}}.
\newblock \bibinfo{journal}{\emph{ICIS 2024 Proceedings}} (\bibinfo{date}{12} \bibinfo{year}{2024}).
\newblock
\urldef\tempurl%
\url{https://aisel.aisnet.org/icis2024/user_behav/user_behav/4}
\showURL{%
\tempurl}


\bibitem[Caliskan et~al\mbox{.}(2017)]%
        {Caliskan2017SemanticsBiases}
\bibfield{author}{\bibinfo{person}{Aylin Caliskan}, \bibinfo{person}{Joanna~J. Bryson}, {and} \bibinfo{person}{Arvind Narayanan}.} \bibinfo{year}{2017}\natexlab{}.
\newblock \showarticletitle{{Semantics derived automatically from language corpora contain human-like biases}}.
\newblock \bibinfo{journal}{\emph{Science}} \bibinfo{volume}{356}, \bibinfo{number}{6334} (\bibinfo{date}{4} \bibinfo{year}{2017}), \bibinfo{pages}{183--186}.
\newblock
\showISSN{10959203}
\href{https://doi.org/10.1126/SCIENCE.AAL4230}{doi:\nolinkurl{10.1126/SCIENCE.AAL4230}}


\bibitem[Chen et~al\mbox{.}(2024)]%
        {Chen2024WhenOpportunities}
\bibfield{author}{\bibinfo{person}{Jin Chen}, \bibinfo{person}{Zheng Liu}, \bibinfo{person}{Xu Huang}, \bibinfo{person}{Chenwang Wu}, \bibinfo{person}{Qi Liu}, \bibinfo{person}{Gangwei Jiang}, \bibinfo{person}{Yuanhao Pu}, \bibinfo{person}{Yuxuan Lei}, \bibinfo{person}{Xiaolong Chen}, \bibinfo{person}{Xingmei Wang}, \bibinfo{person}{Kai Zheng}, \bibinfo{person}{Defu Lian}, {and} \bibinfo{person}{Enhong Chen}.} \bibinfo{year}{2024}\natexlab{}.
\newblock \showarticletitle{{When large language models meet personalization: perspectives of challenges and opportunities}}.
\newblock \bibinfo{journal}{\emph{World Wide Web 2024 27:4}} \bibinfo{volume}{27}, \bibinfo{number}{4} (\bibinfo{date}{6} \bibinfo{year}{2024}), \bibinfo{pages}{1--45}.
\newblock
\showISSN{1573-1413}
\href{https://doi.org/10.1007/S11280-024-01276-1}{doi:\nolinkurl{10.1007/S11280-024-01276-1}}


\bibitem[Chen et~al\mbox{.}(2023)]%
        {Chen2023WouldModel}
\bibfield{author}{\bibinfo{person}{Qian Chen}, \bibinfo{person}{Changqin Yin}, {and} \bibinfo{person}{Yeming Gong}.} \bibinfo{year}{2023}\natexlab{}.
\newblock \showarticletitle{{Would an AI chatbot persuade you: an empirical answer from the elaboration likelihood model}}.
\newblock \bibinfo{journal}{\emph{Information Technology and People}} (\bibinfo{year}{2023}).
\newblock
\showISSN{09593845}
\href{https://doi.org/10.1108/ITP-10-2021-0764}{doi:\nolinkurl{10.1108/ITP-10-2021-0764}}


\bibitem[Cho et~al\mbox{.}(2024)]%
        {Cho2024LLM-BasedSimulations}
\bibfield{author}{\bibinfo{person}{Suhyun Cho}, \bibinfo{person}{Jaeyun Kim}, {and} \bibinfo{person}{Jang~Hyun Kim}.} \bibinfo{year}{2024}\natexlab{}.
\newblock \showarticletitle{{LLM-Based Doppelg{\"{a}}nger Models: Leveraging Synthetic Data for Human-like Responses in Survey Simulations}}.
\newblock \bibinfo{journal}{\emph{IEEE Access}} (\bibinfo{year}{2024}).
\newblock
\showISSN{21693536}
\href{https://doi.org/10.1109/ACCESS.2024.3502219}{doi:\nolinkurl{10.1109/ACCESS.2024.3502219}}


\bibitem[Cialdini and Jacobson(2021)]%
        {Cialdini2021InfluencesBehaviors}
\bibfield{author}{\bibinfo{person}{Robert~B. Cialdini} {and} \bibinfo{person}{Ryan~P. Jacobson}.} \bibinfo{year}{2021}\natexlab{}.
\newblock \showarticletitle{{Influences of social norms on climate change-related behaviors}}.
\newblock \bibinfo{journal}{\emph{Current Opinion in Behavioral Sciences}}  \bibinfo{volume}{42} (\bibinfo{date}{12} \bibinfo{year}{2021}), \bibinfo{pages}{1--8}.
\newblock
\showISSN{2352-1546}
\href{https://doi.org/10.1016/J.COBEHA.2021.01.005}{doi:\nolinkurl{10.1016/J.COBEHA.2021.01.005}}


\bibitem[{CloudResearch}(2025)]%
        {CloudResearch2025ConnectParticipants}
\bibfield{author}{\bibinfo{person}{{CloudResearch}}.} \bibinfo{year}{2025}\natexlab{}.
\newblock \bibinfo{title}{{Connect for Participants}}.
\newblock
\urldef\tempurl%
\url{https://www.cloudresearch.com/products/connect-for-participants/}
\showURL{%
\tempurl}


\bibitem[Costello et~al\mbox{.}(2024)]%
        {Costello2024DurablyAI}
\bibfield{author}{\bibinfo{person}{Thomas~H Costello}, \bibinfo{person}{Gordon Pennycook}, {and} \bibinfo{person}{David~G Rand}.} \bibinfo{year}{2024}\natexlab{}.
\newblock \showarticletitle{{Durably reducing conspiracy beliefs through dialogues with AI}}.
\newblock \bibinfo{journal}{\emph{Science (New York, N.Y.)}} \bibinfo{volume}{385}, \bibinfo{number}{6714} (\bibinfo{year}{2024}), \bibinfo{pages}{eadq1814}.
\newblock
\showISSN{0036-8075}
\href{https://doi.org/10.1126/science.adq1814}{doi:\nolinkurl{10.1126/science.adq1814}}


\bibitem[Demarque et~al\mbox{.}(2015)]%
        {Demarque2015NudgingEnvironment}
\bibfield{author}{\bibinfo{person}{Christophe Demarque}, \bibinfo{person}{Laetitia Charalambides}, \bibinfo{person}{Denis~J. Hilton}, {and} \bibinfo{person}{Laurent Waroquier}.} \bibinfo{year}{2015}\natexlab{}.
\newblock \showarticletitle{{Nudging sustainable consumption: The use of descriptive norms to promote a minority behavior in a realistic online shopping environment}}.
\newblock \bibinfo{journal}{\emph{Journal of Environmental Psychology}}  \bibinfo{volume}{43} (\bibinfo{date}{9} \bibinfo{year}{2015}), \bibinfo{pages}{166--174}.
\newblock
\showISSN{0272-4944}
\href{https://doi.org/10.1016/J.JENVP.2015.06.008}{doi:\nolinkurl{10.1016/J.JENVP.2015.06.008}}


\bibitem[Dickinson et~al\mbox{.}(2016)]%
        {Dickinson2016WhichUSA}
\bibfield{author}{\bibinfo{person}{Janis~L. Dickinson}, \bibinfo{person}{Poppy McLeod}, \bibinfo{person}{Robert Bloomfield}, {and} \bibinfo{person}{Shorna Allred}.} \bibinfo{year}{2016}\natexlab{}.
\newblock \showarticletitle{{Which Moral Foundations Predict Willingness to Make Lifestyle Changes to Avert Climate Change in the USA?}}
\newblock \bibinfo{journal}{\emph{PLOS ONE}} \bibinfo{volume}{11}, \bibinfo{number}{10} (\bibinfo{date}{10} \bibinfo{year}{2016}), \bibinfo{pages}{e0163852}.
\newblock
\showISSN{1932-6203}
\href{https://doi.org/10.1371/JOURNAL.PONE.0163852}{doi:\nolinkurl{10.1371/JOURNAL.PONE.0163852}}


\bibitem[Duarte et~al\mbox{.}(2024)]%
        {Duarte2024EnhancingInsights}
\bibfield{author}{\bibinfo{person}{Paulo Duarte}, \bibinfo{person}{Susana~C. Silva}, \bibinfo{person}{Afonso~S. Roza}, {and} \bibinfo{person}{Joana~Carmo Dias}.} \bibinfo{year}{2024}\natexlab{}.
\newblock \showarticletitle{{Enhancing consumer purchase intentions for sustainable packaging products: An in-depth analysis of key determinants and strategic insights}}.
\newblock \bibinfo{journal}{\emph{Sustainable Futures}}  \bibinfo{volume}{7} (\bibinfo{date}{6} \bibinfo{year}{2024}), \bibinfo{pages}{100193}.
\newblock
\showISSN{2666-1888}
\href{https://doi.org/10.1016/J.SFTR.2024.100193}{doi:\nolinkurl{10.1016/J.SFTR.2024.100193}}


\bibitem[d’Aramon et~al\mbox{.}(2024)]%
        {dAramon2024AssessingChatGPT}
\bibfield{author}{\bibinfo{person}{Ithier d’Aramon}, \bibinfo{person}{Boris Ruf}, {and} \bibinfo{person}{Marcin Detyniecki}.} \bibinfo{year}{2024}\natexlab{}.
\newblock \showarticletitle{{Assessing Carbon Footprint Estimations of ChatGPT}}.
\newblock  (\bibinfo{year}{2024}), \bibinfo{pages}{127--133}.
\newblock
\showISSN{18653537}
\href{https://doi.org/10.1007/978-3-031-59005-4{\_}15}{doi:\nolinkurl{10.1007/978-3-031-59005-4{\_}15}}


\bibitem[Erdfelder et~al\mbox{.}(2009)]%
        {Erdfelder2009StatisticalAnalyses}
\bibfield{author}{\bibinfo{person}{Edgar Erdfelder}, \bibinfo{person}{Franz FAul}, \bibinfo{person}{Axel Buchner}, {and} \bibinfo{person}{Albert~Georg Lang}.} \bibinfo{year}{2009}\natexlab{}.
\newblock \showarticletitle{{Statistical power analyses using G*Power 3.1: Tests for correlation and regression analyses}}.
\newblock \bibinfo{journal}{\emph{Behavior Research Methods}} \bibinfo{volume}{41}, \bibinfo{number}{4} (\bibinfo{year}{2009}), \bibinfo{pages}{1149--1160}.
\newblock
\showISSN{1554351X}
\href{https://doi.org/10.3758/BRM.41.4.1149}{doi:\nolinkurl{10.3758/BRM.41.4.1149}}


\bibitem[Fogg(2003)]%
        {Fogg2003PersuasiveDo}
\bibfield{author}{\bibinfo{person}{B~J Fogg}.} \bibinfo{year}{2003}\natexlab{}.
\newblock \bibinfo{booktitle}{\emph{{Persuasive technology: using computers to change what we think and do}}}.
\newblock \bibinfo{publisher}{Morgan Kaufmann Publishers}, \bibinfo{address}{Amsterdam Boston}.
\newblock
\showISBNx{978-1-55860-643-2}
\href{https://doi.org/10.1016/B978-1-55860-643-2.X5000-8}{doi:\nolinkurl{10.1016/B978-1-55860-643-2.X5000-8}}


\bibitem[Gifford(2011)]%
        {Gifford2011TheAdaptation}
\bibfield{author}{\bibinfo{person}{Robert Gifford}.} \bibinfo{year}{2011}\natexlab{}.
\newblock \showarticletitle{{The dragons of inaction: psychological barriers that limit climate change mitigation and adaptation}}.
\newblock \bibinfo{journal}{\emph{The American psychologist}} \bibinfo{volume}{66}, \bibinfo{number}{4} (\bibinfo{date}{5} \bibinfo{year}{2011}), \bibinfo{pages}{290--302}.
\newblock
\showISSN{1935-990X}
\href{https://doi.org/10.1037/A0023566}{doi:\nolinkurl{10.1037/A0023566}}


\bibitem[Gifford et~al\mbox{.}(2011)]%
        {Gifford2011BehavioralInterventions}
\bibfield{author}{\bibinfo{person}{Robert Gifford}, \bibinfo{person}{Christine Kormos}, {and} \bibinfo{person}{Amanda McIntyre}.} \bibinfo{year}{2011}\natexlab{}.
\newblock \showarticletitle{{Behavioral dimensions of climate change: drivers, responses, barriers, and interventions}}.
\newblock \bibinfo{journal}{\emph{Wiley Interdisciplinary Reviews: Climate Change}} \bibinfo{volume}{2}, \bibinfo{number}{6} (\bibinfo{date}{11} \bibinfo{year}{2011}), \bibinfo{pages}{801--827}.
\newblock
\showISSN{1757-7799}
\href{https://doi.org/10.1002/WCC.143}{doi:\nolinkurl{10.1002/WCC.143}}


\bibitem[Gifford and Nilsson(2014)]%
        {Gifford2014PersonalReview}
\bibfield{author}{\bibinfo{person}{Robert Gifford} {and} \bibinfo{person}{Andreas Nilsson}.} \bibinfo{year}{2014}\natexlab{}.
\newblock \showarticletitle{{Personal and social factors that influence pro-environmental concern and behaviour: a review}}.
\newblock \bibinfo{journal}{\emph{International journal of psychology : Journal international de psychologie}} \bibinfo{volume}{49}, \bibinfo{number}{3} (\bibinfo{year}{2014}), \bibinfo{pages}{141--157}.
\newblock
\showISSN{1464-066X}
\href{https://doi.org/10.1002/IJOP.12034}{doi:\nolinkurl{10.1002/IJOP.12034}}


\bibitem[Giudici(2024)]%
        {Giudici2024PersuasiveTechnology}
\bibfield{author}{\bibinfo{person}{Mathyas Giudici}.} \bibinfo{year}{2024}\natexlab{}.
\newblock \emph{\bibinfo{title}{{Persuasive Conversational Agents to Foster Sustainable Behaviours: Design, Evaluation, and Technology}}}.
\newblock \bibinfo{thesistype}{Ph.\,D. Dissertation}. \bibinfo{school}{Politecnico di Milano, Department of Electronics, Information and Bioengineering}.
\newblock
\href{https://doi.org/10.13140/RG.2.2.32928.70401}{doi:\nolinkurl{10.13140/RG.2.2.32928.70401}}


\bibitem[Goldstein et~al\mbox{.}(2008)]%
        {Goldstein2008AHotels.}
\bibfield{author}{\bibinfo{person}{Noah~J Goldstein}, \bibinfo{person}{Robert~B Cialdini}, {and} \bibinfo{person}{Vladas Griskevicius}.} \bibinfo{year}{2008}\natexlab{}.
\newblock \showarticletitle{{A room with a viewpoint: Using social norms to motivate environmental conservation in hotels.}}
\newblock \bibinfo{journal}{\emph{Journal of Consumer Research}} \bibinfo{volume}{35}, \bibinfo{number}{3} (\bibinfo{year}{2008}), \bibinfo{pages}{472--482}.
\newblock
\showISSN{1537-5277(Electronic),0093-5301(Print)}
\href{https://doi.org/10.1086/586910}{doi:\nolinkurl{10.1086/586910}}


\bibitem[H{\"{a}}m{\"{a}}l{\"{a}}inen et~al\mbox{.}(2023)]%
        {SyntheticStudy2023}
\bibfield{author}{\bibinfo{person}{Perttu H{\"{a}}m{\"{a}}l{\"{a}}inen}, \bibinfo{person}{Mikke Tavast}, {and} \bibinfo{person}{Anton Kunnari}.} \bibinfo{year}{2023}\natexlab{}.
\newblock \showarticletitle{{Evaluating Large Language Models in Generating Synthetic HCI Research Data: a Case Study}}.
\newblock \bibinfo{journal}{\emph{Conference on Human Factors in Computing Systems - Proceedings}} (\bibinfo{date}{4} \bibinfo{year}{2023}), \bibinfo{pages}{19}.
\newblock
\showISBNx{9781450394215}
\href{https://doi.org/10.1145/3544548.3580688}{doi:\nolinkurl{10.1145/3544548.3580688}}


\bibitem[Haws et~al\mbox{.}(2014)]%
        {Haws2014SeeingProducts}
\bibfield{author}{\bibinfo{person}{Kelly~L. Haws}, \bibinfo{person}{Karen~Page Winterich}, {and} \bibinfo{person}{Rebecca~Walker Naylor}.} \bibinfo{year}{2014}\natexlab{}.
\newblock \showarticletitle{{Seeing the world through GREEN-tinted glasses: Green consumption values and responses to environmentally friendly products}}.
\newblock \bibinfo{journal}{\emph{Journal of Consumer Psychology}} \bibinfo{volume}{24}, \bibinfo{number}{3} (\bibinfo{date}{7} \bibinfo{year}{2014}), \bibinfo{pages}{336--354}.
\newblock
\showISSN{1532-7663}
\href{https://doi.org/10.1016/J.JCPS.2013.11.002}{doi:\nolinkurl{10.1016/J.JCPS.2013.11.002}}


\bibitem[Haynes et~al\mbox{.}(2009)]%
        {Haynes2009APopulation}
\bibfield{author}{\bibinfo{person}{Alex~B. Haynes}, \bibinfo{person}{Thomas~G. Weiser}, \bibinfo{person}{William~R. Berry}, \bibinfo{person}{Stuart~R. Lipsitz}, \bibinfo{person}{Abdel-Hadi~S. Breizat}, \bibinfo{person}{E.~Patchen Dellinger}, \bibinfo{person}{Teodoro Herbosa}, \bibinfo{person}{Sudhir Joseph}, \bibinfo{person}{Pascience~L. Kibatala}, \bibinfo{person}{Marie Carmela~M. Lapitan}, \bibinfo{person}{Alan~F. Merry}, \bibinfo{person}{Krishna Moorthy}, \bibinfo{person}{Richard~K. Reznick}, \bibinfo{person}{Bryce Taylor}, {and} \bibinfo{person}{Atul~A. Gawande}.} \bibinfo{year}{2009}\natexlab{}.
\newblock \showarticletitle{{A Surgical Safety Checklist to Reduce Morbidity and Mortality in a Global Population}}.
\newblock \bibinfo{journal}{\emph{New England Journal of Medicine}} \bibinfo{volume}{360}, \bibinfo{number}{5} (\bibinfo{date}{1} \bibinfo{year}{2009}), \bibinfo{pages}{491--499}.
\newblock
\showISSN{0028-4793}
\href{https://doi.org/10.1056/NEJMSA0810119}{doi:\nolinkurl{10.1056/NEJMSA0810119}}


\bibitem[Hestres(2018)]%
        {Hestres2018TakeAdvocacy}
\bibfield{author}{\bibinfo{person}{Luis~E. Hestres}.} \bibinfo{year}{2018}\natexlab{}.
\newblock \showarticletitle{{Take Action Now: Motivational Framing and Action Requests in Climate Advocacy}}.
\newblock \bibinfo{journal}{\emph{Environmental Communication}} \bibinfo{volume}{12}, \bibinfo{number}{4} (\bibinfo{date}{5} \bibinfo{year}{2018}), \bibinfo{pages}{462--479}.
\newblock
\showISSN{17524040}
\href{https://doi.org/10.1080/17524032.2018.1424010}{doi:\nolinkurl{10.1080/17524032.2018.1424010}}


\bibitem[Hewitt et~al\mbox{.}(2024)]%
        {Hewitt2024PredictingModels}
\bibfield{author}{\bibinfo{person}{Luke Hewitt}, \bibinfo{person}{Ashwini Ashokkumar}, \bibinfo{person}{Isaias Ghezae}, {and} \bibinfo{person}{Robb Willer}.} \bibinfo{year}{2024}\natexlab{}.
\newblock \showarticletitle{{Predicting Results of Social Science Experiments Using Large Language Models}}.
\newblock \bibinfo{journal}{\emph{Preprint}} (\bibinfo{date}{8} \bibinfo{year}{2024}).
\newblock
\urldef\tempurl%
\url{https://samim.io/dl/Predicting%20results%20of%20social%20science%20experiments%20using%20large%20language%20models.pdf}
\showURL{%
\tempurl}


\bibitem[Hillebrand and Johannsen(2021)]%
        {Hillebrand2021KlimaKarlSetting}
\bibfield{author}{\bibinfo{person}{Kirsten Hillebrand} {and} \bibinfo{person}{Florian Johannsen}.} \bibinfo{year}{2021}\natexlab{}.
\newblock \showarticletitle{{KlimaKarl – A chabot to promote employees' climate-friendly behavior in an office setting}}.
\newblock \bibinfo{journal}{\emph{SSRN Electronic Journal}} (\bibinfo{year}{2021}).
\newblock
\showISSN{1556-5068}
\href{https://doi.org/10.2139/ssrn.3897674}{doi:\nolinkurl{10.2139/ssrn.3897674}}


\bibitem[Hines et~al\mbox{.}(1987)]%
        {Hines1987AnalysisMeta-Analysis}
\bibfield{author}{\bibinfo{person}{Jody~M. Hines}, \bibinfo{person}{Harold~R. Hungerford}, {and} \bibinfo{person}{Audrey~N. Tomera}.} \bibinfo{year}{1987}\natexlab{}.
\newblock \showarticletitle{{Analysis and Synthesis of Research on Responsible Environmental Behavior: A Meta-Analysis}}.
\newblock \bibinfo{journal}{\emph{The Journal of Environmental Education}} \bibinfo{volume}{18}, \bibinfo{number}{2} (\bibinfo{year}{1987}), \bibinfo{pages}{1--8}.
\newblock
\showISSN{19401892}
\href{https://doi.org/10.1080/00958964.1987.9943482}{doi:\nolinkurl{10.1080/00958964.1987.9943482}}


\bibitem[Jiang et~al\mbox{.}(2024)]%
        {Jiang2024DonaldModels}
\bibfield{author}{\bibinfo{person}{Shapeng Jiang}, \bibinfo{person}{Lijia Wei}, {and} \bibinfo{person}{Chen Zhang}.} \bibinfo{year}{2024}\natexlab{}.
\newblock \showarticletitle{{Donald Trumps in the Virtual Polls: Simulating and Predicting Public Opinions in Surveys Using Large Language Models}}.
\newblock  (\bibinfo{date}{11} \bibinfo{year}{2024}).
\newblock
\href{https://doi.org/10.48550/arXiv.2411.01582}{doi:\nolinkurl{10.48550/arXiv.2411.01582}}


\bibitem[Knutti(2019)]%
        {Knutti2019ClosingChange}
\bibfield{author}{\bibinfo{person}{Reto Knutti}.} \bibinfo{year}{2019}\natexlab{}.
\newblock \showarticletitle{{Closing the Knowledge-Action Gap in Climate Change}}.
\newblock \bibinfo{journal}{\emph{One Earth}} \bibinfo{volume}{1}, \bibinfo{number}{1} (\bibinfo{date}{9} \bibinfo{year}{2019}), \bibinfo{pages}{21--23}.
\newblock
\showISSN{2590-3330}
\href{https://doi.org/10.1016/j.oneear.2019.09.001}{doi:\nolinkurl{10.1016/j.oneear.2019.09.001}}


\bibitem[Kollmuss and Agyeman(2002)]%
        {Kollmuss2002MindBehavior}
\bibfield{author}{\bibinfo{person}{Anja Kollmuss} {and} \bibinfo{person}{Julian Agyeman}.} \bibinfo{year}{2002}\natexlab{}.
\newblock \showarticletitle{{Mind the Gap: Why do people act environmentally and what are the barriers to pro-environmental behavior?}}
\newblock \bibinfo{journal}{\emph{Environmental Education Research}} \bibinfo{volume}{8}, \bibinfo{number}{3} (\bibinfo{year}{2002}), \bibinfo{pages}{239--260}.
\newblock
\showISSN{14695871}
\href{https://doi.org/10.1080/13504620220145401}{doi:\nolinkurl{10.1080/13504620220145401}}


\bibitem[Lau et~al\mbox{.}(2021)]%
        {Lau2021MoralsSciences}
\bibfield{author}{\bibinfo{person}{Jacqueline~D. Lau}, \bibinfo{person}{Andrew~M. Song}, \bibinfo{person}{Tiffany Morrison}, \bibinfo{person}{Michael Fabinyi}, \bibinfo{person}{Katrina Brown}, \bibinfo{person}{Jessica Blythe}, \bibinfo{person}{Edward~H. Allison}, {and} \bibinfo{person}{William~Neil Adger}.} \bibinfo{year}{2021}\natexlab{}.
\newblock \showarticletitle{{Morals and climate decision-making: insights from social and behavioural sciences}}.
\newblock \bibinfo{journal}{\emph{Current Opinion in Environmental Sustainability}}  \bibinfo{volume}{52} (\bibinfo{date}{10} \bibinfo{year}{2021}), \bibinfo{pages}{27--35}.
\newblock
\showISSN{1877-3435}
\href{https://doi.org/10.1016/J.COSUST.2021.06.005}{doi:\nolinkurl{10.1016/J.COSUST.2021.06.005}}


\bibitem[Lehner et~al\mbox{.}(2016)]%
        {Lehner2016NudgingBehaviour}
\bibfield{author}{\bibinfo{person}{Matthias Lehner}, \bibinfo{person}{Oksana Mont}, {and} \bibinfo{person}{Eva Heiskanen}.} \bibinfo{year}{2016}\natexlab{}.
\newblock \showarticletitle{{Nudging – A promising tool for sustainable consumption behaviour?}}
\newblock \bibinfo{journal}{\emph{Journal of Cleaner Production}}  \bibinfo{volume}{134} (\bibinfo{date}{10} \bibinfo{year}{2016}), \bibinfo{pages}{166--177}.
\newblock
\showISSN{0959-6526}
\href{https://doi.org/10.1016/J.JCLEPRO.2015.11.086}{doi:\nolinkurl{10.1016/J.JCLEPRO.2015.11.086}}


\bibitem[Ligozat et~al\mbox{.}(2021)]%
        {Ligozat2021UnravelingEnvironment}
\bibfield{author}{\bibinfo{person}{Anne-Laure Ligozat}, \bibinfo{person}{Julien Lef{\`{e}}vre}, \bibinfo{person}{Aurélie Bugeau}, {and} \bibinfo{person}{Jacques Combaz}.} \bibinfo{year}{2021}\natexlab{}.
\newblock \showarticletitle{{Unraveling the Hidden Environmental Impacts of AI Solutions for Environment}}.
\newblock \bibinfo{journal}{\emph{"Towards the Sustainability of AI; Multi-Disciplinary Approaches to Investigate the Hidden Costs of AI"}} (\bibinfo{date}{10} \bibinfo{year}{2021}).
\newblock
\urldef\tempurl%
\url{https://arxiv.org/abs/2110.11822v2}
\showURL{%
\tempurl}


\bibitem[Lynas et~al\mbox{.}(2021)]%
        {Lynas2021GreaterLiterature}
\bibfield{author}{\bibinfo{person}{Mark Lynas}, \bibinfo{person}{Benjamin~Z. Houlton}, {and} \bibinfo{person}{Simon Perry}.} \bibinfo{year}{2021}\natexlab{}.
\newblock \showarticletitle{{Greater than 99{\%} consensus on human caused climate change in the peer-reviewed scientific literature}}.
\newblock \bibinfo{journal}{\emph{Environmental Research Letters}} \bibinfo{volume}{16}, \bibinfo{number}{11} (\bibinfo{date}{10} \bibinfo{year}{2021}), \bibinfo{pages}{114005}.
\newblock
\showISSN{1748-9326}
\href{https://doi.org/10.1088/1748-9326/AC2966}{doi:\nolinkurl{10.1088/1748-9326/AC2966}}


\bibitem[Matz et~al\mbox{.}(2024)]%
        {Matz2024TheScale}
\bibfield{author}{\bibinfo{person}{S.~C. Matz}, \bibinfo{person}{J.~D. Teeny}, \bibinfo{person}{S.~S. Vaid}, \bibinfo{person}{H. Peters}, \bibinfo{person}{G.~M. Harari}, {and} \bibinfo{person}{M. Cerf}.} \bibinfo{year}{2024}\natexlab{}.
\newblock \showarticletitle{{The potential of generative AI for personalized persuasion at scale}}.
\newblock \bibinfo{journal}{\emph{Scientific Reports 2024 14:1}} \bibinfo{volume}{14}, \bibinfo{number}{1} (\bibinfo{date}{2} \bibinfo{year}{2024}), \bibinfo{pages}{1--16}.
\newblock
\showISBNx{0123456789}
\showISSN{2045-2322}
\href{https://doi.org/10.1038/s41598-024-53755-0}{doi:\nolinkurl{10.1038/s41598-024-53755-0}}


\bibitem[Myers et~al\mbox{.}(2021)]%
        {Myers2021ConsensusLater}
\bibfield{author}{\bibinfo{person}{Krista~F. Myers}, \bibinfo{person}{Peter~T. Doran}, \bibinfo{person}{John Cook}, \bibinfo{person}{John~E. Kotcher}, {and} \bibinfo{person}{Teresa~A. Myers}.} \bibinfo{year}{2021}\natexlab{}.
\newblock \showarticletitle{{Consensus revisited: quantifying scientific agreement on climate change and climate expertise among Earth scientists 10 years later}}.
\newblock \bibinfo{journal}{\emph{Environmental Research Letters}} \bibinfo{volume}{16}, \bibinfo{number}{10} (\bibinfo{date}{10} \bibinfo{year}{2021}), \bibinfo{pages}{104030}.
\newblock
\showISSN{1748-9326}
\href{https://doi.org/10.1088/1748-9326/AC2774}{doi:\nolinkurl{10.1088/1748-9326/AC2774}}


\bibitem[Nguyen et~al\mbox{.}(2019)]%
        {Nguyen2019GreenGap}
\bibfield{author}{\bibinfo{person}{Hung~Vu Nguyen}, \bibinfo{person}{Cuong~Hung Nguyen}, {and} \bibinfo{person}{Thoa Thi~Bao Hoang}.} \bibinfo{year}{2019}\natexlab{}.
\newblock \showarticletitle{{Green consumption: Closing the intention-behavior gap}}.
\newblock \bibinfo{journal}{\emph{Sustainable Development}} \bibinfo{volume}{27}, \bibinfo{number}{1} (\bibinfo{date}{1} \bibinfo{year}{2019}), \bibinfo{pages}{118--129}.
\newblock
\showISSN{1099-1719}
\href{https://doi.org/10.1002/SD.1875}{doi:\nolinkurl{10.1002/SD.1875}}


\bibitem[on~Climate Change~(IPCC)(2023)]%
        {onClimateChangeIPCC2023ClimateChange}
\bibfield{author}{\bibinfo{person}{Intergovernmental~Panel on Climate Change~(IPCC)}.} \bibinfo{year}{2023}\natexlab{}.
\newblock \bibinfo{title}{{Climate Change 2023: Synthesis Report. A Report of the Intergovernmental Panel on Climate Change. Contribution of Working Groups I, II and III to the Sixth Assessment Report of the Intergovernmental Panel on Climate Change}}.
\newblock
\urldef\tempurl%
\url{https://www.ipcc.ch/report/sixth-assessment-report-cycle/}
\showURL{%
\tempurl}


\bibitem[Pataranutaporn et~al\mbox{.}(2024)]%
        {Pataranutaporn2024FutureSelf-Continuity}
\bibfield{author}{\bibinfo{person}{Pat Pataranutaporn}, \bibinfo{person}{Kavin Winson}, \bibinfo{person}{Peggy Yin}, \bibinfo{person}{Auttasak Lapapirojn}, \bibinfo{person}{Pichayoot Ouppaphan}, \bibinfo{person}{Monchai Lertsutthiwong}, \bibinfo{person}{Pattie Maes}, {and} \bibinfo{person}{Hal Hershfield}.} \bibinfo{year}{2024}\natexlab{}.
\newblock \showarticletitle{{Future You: A Conversation with an AI-Generated Future Self Reduces Anxiety, Negative Emotions, and Increases Future Self-Continuity}}.
\newblock  (\bibinfo{year}{2024}).
\newblock
\href{https://doi.org/10.48550/arXiv.2405.12514}{doi:\nolinkurl{10.48550/arXiv.2405.12514}}


\bibitem[Qin et~al\mbox{.}(2024)]%
        {Qin2024TheSelf-efficacy}
\bibfield{author}{\bibinfo{person}{Ziqi Qin}, \bibinfo{person}{Qi Wu}, \bibinfo{person}{Cuihua Bi}, \bibinfo{person}{Yanwei Deng}, {and} \bibinfo{person}{Qiuyun Hu}.} \bibinfo{year}{2024}\natexlab{}.
\newblock \showarticletitle{{The relationship between climate change anxiety and pro-environmental behavior in adolescents: the mediating role of future self-continuity and the moderating role of green self-efficacy}}.
\newblock \bibinfo{journal}{\emph{BMC Psychology}} \bibinfo{volume}{12}, \bibinfo{number}{1} (\bibinfo{date}{12} \bibinfo{year}{2024}), \bibinfo{pages}{1--12}.
\newblock
\showISSN{20507283}
\href{https://doi.org/10.1186/S40359-024-01746-1}{doi:\nolinkurl{10.1186/S40359-024-01746-1}}


\bibitem[Soares et~al\mbox{.}(2021)]%
        {Soares2021PublicBehaviours}
\bibfield{author}{\bibinfo{person}{Joana Soares}, \bibinfo{person}{Isabel Miguel}, \bibinfo{person}{Cátia Ven{\^{a}}ncio}, \bibinfo{person}{Isabel Lopes}, {and} \bibinfo{person}{Miguel Oliveira}.} \bibinfo{year}{2021}\natexlab{}.
\newblock \showarticletitle{{Public views on plastic pollution: Knowledge, perceived impacts, and pro-environmental behaviours}}.
\newblock \bibinfo{journal}{\emph{Journal of Hazardous Materials}}  \bibinfo{volume}{412} (\bibinfo{date}{6} \bibinfo{year}{2021}), \bibinfo{pages}{125227}.
\newblock
\showISSN{0304-3894}
\href{https://doi.org/10.1016/J.JHAZMAT.2021.125227}{doi:\nolinkurl{10.1016/J.JHAZMAT.2021.125227}}


\bibitem[Sparkman et~al\mbox{.}(2021)]%
        {Sparkman2021HowSolution}
\bibfield{author}{\bibinfo{person}{Gregg Sparkman}, \bibinfo{person}{Lauren Howe}, {and} \bibinfo{person}{Greg Walton}.} \bibinfo{year}{2021}\natexlab{}.
\newblock \showarticletitle{{How social norms are often a barrier to addressing climate change but can be part of the solution}}.
\newblock \bibinfo{journal}{\emph{Behavioural Public Policy}} \bibinfo{volume}{5}, \bibinfo{number}{4} (\bibinfo{year}{2021}), \bibinfo{pages}{528--555}.
\newblock
\showISSN{2398-063X}
\href{https://doi.org/DOI: 10.1017/bpp.2020.42}{doi:\nolinkurl{DOI: 10.1017/bpp.2020.42}}


\bibitem[Sparkman and Walton(2017)]%
        {Sparkman2017DynamicCounterinformative}
\bibfield{author}{\bibinfo{person}{Gregg Sparkman} {and} \bibinfo{person}{Gregory~M. Walton}.} \bibinfo{year}{2017}\natexlab{}.
\newblock \showarticletitle{{Dynamic Norms Promote Sustainable Behavior, Even if It Is Counterinformative}}.
\newblock \bibinfo{journal}{\emph{https://doi.org/10.1177/0956797617719950}} \bibinfo{volume}{28}, \bibinfo{number}{11} (\bibinfo{date}{9} \bibinfo{year}{2017}), \bibinfo{pages}{1663--1674}.
\newblock
\showISSN{14679280}
\href{https://doi.org/10.1177/0956797617719950}{doi:\nolinkurl{10.1177/0956797617719950}}


\bibitem[Thaler and Sunstein(2008)]%
        {Thaler2008Nudge:Happiness}
\bibfield{author}{\bibinfo{person}{Richard~H Thaler} {and} \bibinfo{person}{Cass~R Sunstein}.} \bibinfo{year}{2008}\natexlab{}.
\newblock \bibinfo{booktitle}{\emph{{Nudge: improving decisions about health, wealth, and happiness}}}.
\newblock \bibinfo{publisher}{Yale University Press}, \bibinfo{address}{New Haven}. 293 pages.
\newblock
\showISBNx{978-0-300-12223-7}


\bibitem[Tjuatja et~al\mbox{.}(2023)]%
        {Tjuatja2023DoDesign}
\bibfield{author}{\bibinfo{person}{Lindia Tjuatja}, \bibinfo{person}{Valerie Chen}, \bibinfo{person}{Tongshuang Wu}, \bibinfo{person}{Ameet Talwalkwar}, {and} \bibinfo{person}{Graham Neubig}.} \bibinfo{year}{2023}\natexlab{}.
\newblock \showarticletitle{{Do LLMs exhibit human-like response biases? A case study in survey design}}.
\newblock \bibinfo{journal}{\emph{Transactions of the Association for Computational Linguistics}}  \bibinfo{volume}{12} (\bibinfo{date}{11} \bibinfo{year}{2023}), \bibinfo{pages}{1011--1026}.
\newblock
\showISSN{2307387X}
\href{https://doi.org/10.1162/tacl{\_}a{\_}00685}{doi:\nolinkurl{10.1162/tacl{\_}a{\_}00685}}


\bibitem[van Valkengoed et~al\mbox{.}(2021)]%
        {vanValkengoed2021DevelopmentScale}
\bibfield{author}{\bibinfo{person}{A.~M. van Valkengoed}, \bibinfo{person}{L. Steg}, {and} \bibinfo{person}{G. Perlaviciute}.} \bibinfo{year}{2021}\natexlab{}.
\newblock \showarticletitle{{Development and validation of a climate change perceptions scale}}.
\newblock \bibinfo{journal}{\emph{Journal of Environmental Psychology}}  \bibinfo{volume}{76} (\bibinfo{date}{8} \bibinfo{year}{2021}), \bibinfo{pages}{101652}.
\newblock
\showISSN{0272-4944}
\href{https://doi.org/10.1016/J.JENVP.2021.101652}{doi:\nolinkurl{10.1016/J.JENVP.2021.101652}}


\bibitem[Vlasceanu et~al\mbox{.}(2024)]%
        {Vlasceanu2024AddressingCountries}
\bibfield{author}{\bibinfo{person}{Madalina Vlasceanu}, \bibinfo{person}{Kimberly~C Doell}, \bibinfo{person}{Joseph~B Bak-Coleman}, \bibinfo{person}{Boryana Todorova}, \bibinfo{person}{Michael~M Berkebile-Weinberg}, \bibinfo{person}{Samantha~J Grayson}, \bibinfo{person}{Yash Patel}, \bibinfo{person}{Danielle Goldwert}, \bibinfo{person}{Yifei Pei}, \bibinfo{person}{Alek Chakroff}, \bibinfo{person}{Ekaterina Pronizius}, \bibinfo{person}{Karlijn~L van~den Broek}, \bibinfo{person}{Denisa Vlasceanu}, \bibinfo{person}{Sara Constantino}, \bibinfo{person}{Michael~J Morais}, \bibinfo{person}{Philipp Schumann}, \bibinfo{person}{Steve Rathje}, \bibinfo{person}{Ke Fang}, \bibinfo{person}{Salvatore~Maria Aglioti}, \bibinfo{person}{Mark Alfano}, \bibinfo{person}{Andy~J Alvarado-Yepez}, \bibinfo{person}{Angélica Andersen}, \bibinfo{person}{Frederik Anseel}, \bibinfo{person}{Matthew A~J Apps}, \bibinfo{person}{Chillar Asadli}, \bibinfo{person}{Fonda~Jane Awuor}, \bibinfo{person}{Flavio Azevedo}, \bibinfo{person}{Piero
  Basaglia}, \bibinfo{person}{Jocelyn~J B{\'{e}}langer}, \bibinfo{person}{Sebastian Berger}, \bibinfo{person}{Paul Bertin}, \bibinfo{person}{Michał Bia{\l}ek}, \bibinfo{person}{Olga Bialobrzeska}, \bibinfo{person}{Michelle Blaya-Burgo}, \bibinfo{person}{Daniëlle N~M Bleize}, \bibinfo{person}{Simen B{\o}}, \bibinfo{person}{Lea Boecker}, \bibinfo{person}{Paulo~S Boggio}, \bibinfo{person}{Sylvie Borau}, \bibinfo{person}{Björn Bos}, \bibinfo{person}{Ayoub Bouguettaya}, \bibinfo{person}{Markus Brauer}, \bibinfo{person}{Cameron Brick}, \bibinfo{person}{Tymofii Brik}, \bibinfo{person}{Roman Briker}, \bibinfo{person}{Tobias Brosch}, \bibinfo{person}{Ondrej Buchel}, \bibinfo{person}{Daniel Buonauro}, \bibinfo{person}{Radhika Butalia}, \bibinfo{person}{Héctor Carvacho}, \bibinfo{person}{Sarah A~E Chamberlain}, \bibinfo{person}{Hang-Yee Chan}, \bibinfo{person}{Dawn Chow}, \bibinfo{person}{Dongil Chung}, \bibinfo{person}{Luca Cian}, \bibinfo{person}{Noa Cohen-Eick}, \bibinfo{person}{Luis~Sebastian Contreras-Huerta},
  \bibinfo{person}{Davide Contu}, \bibinfo{person}{Vladimir Cristea}, \bibinfo{person}{Jo Cutler}, \bibinfo{person}{Silvana D'Ottone}, \bibinfo{person}{Jonas De~Keersmaecker}, \bibinfo{person}{Sarah Delcourt}, \bibinfo{person}{Sylvain Delouv{\'{e}}e}, \bibinfo{person}{Kathi Diel}, \bibinfo{person}{Benjamin~D Douglas}, \bibinfo{person}{Moritz~A Drupp}, \bibinfo{person}{Shreya Dubey}, \bibinfo{person}{Jānis Ekmanis}, \bibinfo{person}{Christian~T Elbaek}, \bibinfo{person}{Mahmoud Elsherif}, \bibinfo{person}{Iris~M Engelhard}, \bibinfo{person}{Yannik~A Escher}, \bibinfo{person}{Tom~W Etienne}, \bibinfo{person}{Laura Farage}, \bibinfo{person}{Ana~Rita Farias}, \bibinfo{person}{Stefan Feuerriegel}, \bibinfo{person}{Andrej Findor}, \bibinfo{person}{Lucia Freira}, \bibinfo{person}{Malte Friese}, \bibinfo{person}{Neil~Philip Gains}, \bibinfo{person}{Albina Gallyamova}, \bibinfo{person}{Sandra~J Geiger}, \bibinfo{person}{Oliver Genschow}, \bibinfo{person}{Biljana Gjoneska}, \bibinfo{person}{Theofilos Gkinopoulos},
  \bibinfo{person}{Beth Goldberg}, \bibinfo{person}{Amit Goldenberg}, \bibinfo{person}{Sarah Gradidge}, \bibinfo{person}{Simone Grassini}, \bibinfo{person}{Kurt Gray}, \bibinfo{person}{Sonja Grelle}, \bibinfo{person}{Siobhán~M Griffin}, \bibinfo{person}{Lusine Grigoryan}, \bibinfo{person}{Ani Grigoryan}, \bibinfo{person}{Dmitry Grigoryev}, \bibinfo{person}{June Gruber}, \bibinfo{person}{Johnrev Guilaran}, \bibinfo{person}{Britt Hadar}, \bibinfo{person}{Ulf J~J Hahnel}, \bibinfo{person}{Eran Halperin}, \bibinfo{person}{Annelie~J Harvey}, \bibinfo{person}{Christian A~P Haugestad}, \bibinfo{person}{Aleksandra~M Herman}, \bibinfo{person}{Hal~E Hershfield}, \bibinfo{person}{Toshiyuki Himichi}, \bibinfo{person}{Donald~W Hine}, \bibinfo{person}{Wilhelm Hofmann}, \bibinfo{person}{Lauren Howe}, \bibinfo{person}{Enma~T Huaman-Chulluncuy}, \bibinfo{person}{Guanxiong Huang}, \bibinfo{person}{Tatsunori Ishii}, \bibinfo{person}{Ayahito Ito}, \bibinfo{person}{Fanli Jia}, \bibinfo{person}{John~T Jost},
  \bibinfo{person}{Veljko Jovanovi{\'{c}}}, \bibinfo{person}{Dominika Jurgiel}, \bibinfo{person}{Ondřej K{\'{a}}cha}, \bibinfo{person}{Reeta Kankaanp{\"{a}}{\"{a}}}, \bibinfo{person}{Jaroslaw Kantorowicz}, \bibinfo{person}{Elena Kantorowicz-Reznichenko}, \bibinfo{person}{Keren Kaplan~Mintz}, \bibinfo{person}{Ilker Kaya}, \bibinfo{person}{Ozgur Kaya}, \bibinfo{person}{Narine Khachatryan}, \bibinfo{person}{Anna Klas}, \bibinfo{person}{Colin Klein}, \bibinfo{person}{Christian~A Kl{\"{o}}ckner}, \bibinfo{person}{Lina Koppel}, \bibinfo{person}{Alexandra~I Kosachenko}, \bibinfo{person}{Emily~J Kothe}, \bibinfo{person}{Ruth Krebs}, \bibinfo{person}{Amy~R Krosch}, \bibinfo{person}{Andre P~M Krouwel}, \bibinfo{person}{Yara Kyrychenko}, \bibinfo{person}{Maria Lagomarsino}, \bibinfo{person}{Claus Lamm}, \bibinfo{person}{Florian Lange}, \bibinfo{person}{Julia Lee~Cunningham}, \bibinfo{person}{Jeffrey Lees}, \bibinfo{person}{Tak~Yan Leung}, \bibinfo{person}{Neil Levy}, \bibinfo{person}{Patricia~L Lockwood},
  \bibinfo{person}{Chiara Longoni}, \bibinfo{person}{Alberto L{\'{o}}pez~Ortega}, \bibinfo{person}{David~D Loschelder}, \bibinfo{person}{Jackson~G Lu}, \bibinfo{person}{Yu Luo}, \bibinfo{person}{Joseph Luomba}, \bibinfo{person}{Annika~E Lutz}, \bibinfo{person}{Johann~M Majer}, \bibinfo{person}{Ezra Markowitz}, \bibinfo{person}{Abigail~A Marsh}, \bibinfo{person}{Karen~Louise Mascarenhas}, \bibinfo{person}{Bwambale Mbilingi}, \bibinfo{person}{Winfred Mbungu}, \bibinfo{person}{Cillian McHugh}, \bibinfo{person}{Marijn H~C Meijers}, \bibinfo{person}{Hugo Mercier}, \bibinfo{person}{Fenant~Laurent Mhagama}, \bibinfo{person}{Katerina Michalakis}, \bibinfo{person}{Nace Mikus}, \bibinfo{person}{Sarah Milliron}, \bibinfo{person}{Panagiotis Mitkidis}, \bibinfo{person}{Fredy~S Monge-Rodr{\'{i}}guez}, \bibinfo{person}{Youri~L Mora}, \bibinfo{person}{David Moreau}, \bibinfo{person}{Kosuke Motoki}, \bibinfo{person}{Manuel Moyano}, \bibinfo{person}{Mathilde Mus}, \bibinfo{person}{Joaquin Navajas}, \bibinfo{person}{Tam~Luong
  Nguyen}, \bibinfo{person}{Dung~Minh Nguyen}, \bibinfo{person}{Trieu Nguyen}, \bibinfo{person}{Laura Niemi}, \bibinfo{person}{Sari R~R Nijssen}, \bibinfo{person}{Gustav Nilsonne}, \bibinfo{person}{Jonas~P Nitschke}, \bibinfo{person}{Laila Nockur}, \bibinfo{person}{Ritah Okura}, \bibinfo{person}{Sezin {\"{O}}ner}, \bibinfo{person}{Asil~Ali {\"{O}}zdo{\u{g}}ru}, \bibinfo{person}{Helena Palumbo}, \bibinfo{person}{Costas Panagopoulos}, \bibinfo{person}{Maria~Serena Panasiti}, \bibinfo{person}{Philip P{\"{a}}rnamets}, \bibinfo{person}{Mariola Paruzel-Czachura}, \bibinfo{person}{Yuri~G Pavlov}, \bibinfo{person}{César Pay{\'{a}}n-G{\'{o}}mez}, \bibinfo{person}{Adam~R Pearson}, \bibinfo{person}{Leonor Pereira~da Costa}, \bibinfo{person}{Hannes~M Petrowsky}, \bibinfo{person}{Stefan Pfattheicher}, \bibinfo{person}{Nhat~Tan Pham}, \bibinfo{person}{Vladimir Ponizovskiy}, \bibinfo{person}{Clara Pretus}, \bibinfo{person}{Gabriel~G R{\^{e}}go}, \bibinfo{person}{Ritsaart Reimann}, \bibinfo{person}{Shawn~A Rhoads},
  \bibinfo{person}{Julian Riano-Moreno}, \bibinfo{person}{Isabell Richter}, \bibinfo{person}{Jan~Philipp R{\"{o}}er}, \bibinfo{person}{Jahred Rosa-Sullivan}, \bibinfo{person}{Robert~M Ross}, \bibinfo{person}{Anandita Sabherwal}, \bibinfo{person}{Toshiki Saito}, \bibinfo{person}{Oriane Sarrasin}, \bibinfo{person}{Nicolas Say}, \bibinfo{person}{Katharina Schmid}, \bibinfo{person}{Michael~T Schmitt}, \bibinfo{person}{Philipp Schoenegger}, \bibinfo{person}{Christin Scholz}, \bibinfo{person}{Mariah~G Schug}, \bibinfo{person}{Stefan Schulreich}, \bibinfo{person}{Ganga Shreedhar}, \bibinfo{person}{Eric Shuman}, \bibinfo{person}{Smadar Sivan}, \bibinfo{person}{Hallgeir Sj{\aa}stad}, \bibinfo{person}{Meikel Soliman}, \bibinfo{person}{Katia Soud}, \bibinfo{person}{Tobia Spampatti}, \bibinfo{person}{Gregg Sparkman}, \bibinfo{person}{Ognen Spasovski}, \bibinfo{person}{Samantha~K Stanley}, \bibinfo{person}{Jessica~A Stern}, \bibinfo{person}{Noel Strahm}, \bibinfo{person}{Yasushi Suko}, \bibinfo{person}{Sunhae Sul},
  \bibinfo{person}{Stylianos Syropoulos}, \bibinfo{person}{Neil~C Taylor}, \bibinfo{person}{Elisa Tedaldi}, \bibinfo{person}{Gustav Tingh{\"{o}}g}, \bibinfo{person}{Luu Duc~Toan Huynh}, \bibinfo{person}{Giovanni~Antonio Travaglino}, \bibinfo{person}{Manos Tsakiris}, \bibinfo{person}{İlayda T{\"{u}}ter}, \bibinfo{person}{Michael Tyrala}, \bibinfo{person}{Özden~Melis Ulu{\u{g}}}, \bibinfo{person}{Arkadiusz Urbanek}, \bibinfo{person}{Danila Valko}, \bibinfo{person}{Sander van~der Linden}, \bibinfo{person}{Kevin van Schie}, \bibinfo{person}{Aart van Stekelenburg}, \bibinfo{person}{Edmunds Vanags}, \bibinfo{person}{Daniel V{\"{a}}stfj{\"{a}}ll}, \bibinfo{person}{Stepan Vesely}, \bibinfo{person}{Jáchym Vintr}, \bibinfo{person}{Marek Vranka}, \bibinfo{person}{Patrick~Otuo Wanguche}, \bibinfo{person}{Robb Willer}, \bibinfo{person}{Adrian~Dominik Wojcik}, \bibinfo{person}{Rachel Xu}, \bibinfo{person}{Anjali Yadav}, \bibinfo{person}{Magdalena Zawisza}, \bibinfo{person}{Xian Zhao}, \bibinfo{person}{Jiaying Zhao},
  \bibinfo{person}{Dawid {\.{Z}}uk}, {and} \bibinfo{person}{Jay~J Van~Bavel}.} \bibinfo{year}{2024}\natexlab{}.
\newblock \showarticletitle{{Addressing climate change with behavioral science: A global intervention tournament in 63 countries}}.
\newblock \bibinfo{journal}{\emph{Science Advances}} \bibinfo{volume}{10}, \bibinfo{number}{6} (\bibinfo{date}{4} \bibinfo{year}{2024}), \bibinfo{pages}{eadj5778}.
\newblock
\href{https://doi.org/10.1126/sciadv.adj5778}{doi:\nolinkurl{10.1126/sciadv.adj5778}}


\bibitem[Wan et~al\mbox{.}(2023)]%
        {Wan2023KellyLetters}
\bibfield{author}{\bibinfo{person}{Yixin Wan}, \bibinfo{person}{George Pu}, \bibinfo{person}{Jiao Sun}, \bibinfo{person}{Aparna Garimella}, \bibinfo{person}{Kai~Wei Chang}, {and} \bibinfo{person}{Nanyun Peng}.} \bibinfo{year}{2023}\natexlab{}.
\newblock \showarticletitle{{"Kelly is a Warm Person, Joseph is a Role Model": Gender Biases in LLM-Generated Reference Letters}}.
\newblock \bibinfo{journal}{\emph{Findings of the Association for Computational Linguistics: EMNLP 2023}} (\bibinfo{date}{10} \bibinfo{year}{2023}), \bibinfo{pages}{3730--3748}.
\newblock
\showISBNx{9798891760615}
\href{https://doi.org/10.18653/v1/2023.findings-emnlp.243}{doi:\nolinkurl{10.18653/v1/2023.findings-emnlp.243}}


\bibitem[Wang et~al\mbox{.}(2023)]%
        {Wang2023ExploringSuccess}
\bibfield{author}{\bibinfo{person}{Ting Wang}, \bibinfo{person}{Brady~D. Lund}, \bibinfo{person}{Agostino Marengo}, \bibinfo{person}{Alessandro Pagano}, \bibinfo{person}{Nishith~Reddy Mannuru}, \bibinfo{person}{Zoë~A. Teel}, {and} \bibinfo{person}{Jenny Pange}.} \bibinfo{year}{2023}\natexlab{}.
\newblock \showarticletitle{{Exploring the Potential Impact of Artificial Intelligence (AI) on International Students in Higher Education: Generative AI, Chatbots, Analytics, and International Student Success}}.
\newblock \bibinfo{journal}{\emph{Applied Sciences 2023, Vol. 13, Page 6716}} \bibinfo{volume}{13}, \bibinfo{number}{11} (\bibinfo{date}{5} \bibinfo{year}{2023}), \bibinfo{pages}{6716}.
\newblock
\showISSN{2076-3417}
\href{https://doi.org/10.3390/APP13116716}{doi:\nolinkurl{10.3390/APP13116716}}


\bibitem[Zhao et~al\mbox{.}(2017)]%
        {Zhao2017MenConstraints}
\bibfield{author}{\bibinfo{person}{Jieyu Zhao}, \bibinfo{person}{Tianlu Wang}, \bibinfo{person}{Mark Yatskar}, \bibinfo{person}{Vicente Ordonez}, {and} \bibinfo{person}{Kai~Wei Chang}.} \bibinfo{year}{2017}\natexlab{}.
\newblock \showarticletitle{{Men Also Like Shopping: Reducing Gender Bias Amplification using Corpus-level Constraints}}.
\newblock \bibinfo{journal}{\emph{EMNLP 2017 - Conference on Empirical Methods in Natural Language Processing, Proceedings}} (\bibinfo{year}{2017}), \bibinfo{pages}{2979--2989}.
\newblock
\showISBNx{9781945626838}
\href{https://doi.org/10.18653/V1/D17-1323}{doi:\nolinkurl{10.18653/V1/D17-1323}}


\bibitem[Zhong et~al\mbox{.}(2024)]%
        {Zhong2024MemoryBank:Memory}
\bibfield{author}{\bibinfo{person}{Wanjun Zhong}, \bibinfo{person}{Lianghong Guo}, \bibinfo{person}{Qiqi Gao}, \bibinfo{person}{He Ye}, {and} \bibinfo{person}{Yanlin Wang}.} \bibinfo{year}{2024}\natexlab{}.
\newblock \showarticletitle{{MemoryBank: Enhancing Large Language Models with Long-Term Memory}}.
\newblock \bibinfo{journal}{\emph{Proceedings of the AAAI Conference on Artificial Intelligence}} \bibinfo{volume}{38}, \bibinfo{number}{17} (\bibinfo{date}{3} \bibinfo{year}{2024}), \bibinfo{pages}{19724--19731}.
\newblock
\showISSN{2374-3468}
\href{https://doi.org/10.1609/AAAI.V38I17.29946}{doi:\nolinkurl{10.1609/AAAI.V38I17.29946}}


\end{thebibliography}
\end{document}